\documentclass[sigconf, nonacm]{acmart}

\usepackage{graphicx}
\usepackage{subfig}
\usepackage{amsmath}
\usepackage{nicefrac}
\usepackage{listings}
\usepackage{appendix}
\usepackage{xspace}
\usepackage{etoolbox}
\usepackage{booktabs} 
\usepackage[utf8]{inputenc}
\usepackage{xcolor}
\usepackage{verbatim}
\usepackage[subtle,title=tight]{savetrees}
\PassOptionsToPackage{hyphens}{url}
\usepackage{hyperref}
\usepackage{multirow}
\usepackage{colortbl}
\usepackage{microtype}
\usepackage{balance}
\usepackage{algorithm,algorithmicx,algpseudocode}

\newcommand{\tsunami}{Tsunami\xspace}
\newcommand{\augmentedgrid}{Augmented Grid\xspace}
\newcommand{\augmentedgrids}{Augmented Grids\xspace}

\newcommand{\gridtree}{Grid Tree\xspace}

\newcommand{\NewPara}[1]{\vspace{4pt}\noindent{\bf #1}}
\newcommand{\Section}[1]{\S\ref{sec:#1}}

\newcommand{\Figure}[1]{Fig.~\ref{fig:#1}}
\newcommand{\Table}[1]{Tab.~\ref{tab:#1}}

\def\compactify{\itemsep=0pt \topsep=0pt \partopsep=0pt \parsep=0pt \leftmargin=0.5cm}
\let\latexusecounter=\usecounter

\newenvironment{CompactEnumerate} 
  {\def\usecounter{\compactify\latexusecounter} 
   \begin{enumerate}}
  {\end{enumerate}\let\usecounter=\latexusecounter}

\DeclareMathOperator*{\argmax}{arg\,max}

\newcounter{todocounter}

\begin{document}
\fancyhead{}
\title{Tsunami: A Learned Multi-dimensional Index for Correlated Data and Skewed Workloads}

\author{Jialin Ding, Vikram Nathan, Mohammad Alizadeh, Tim Kraska}
\affiliation{Massachusetts Insititute of Technology
}

\begin{abstract}
Filtering data based on predicates is one of the most fundamental operations for any modern data warehouse. Techniques to accelerate the execution of filter expressions include clustered indexes, specialized sort orders (e.g., Z-order), multi-dimensional indexes, and, for high selectivity queries, secondary indexes. However, these schemes are hard to tune and their performance is inconsistent. Recent work on learned multi-dimensional indexes has introduced the idea of automatically optimizing an index for a particular dataset and workload. However, the performance of that work suffers in the presence of correlated data and skewed query workloads, both of which are common in real applications.
In this paper, we introduce \tsunami, which addresses these limitations to achieve up to 6$\times$ faster query performance and up to 8$\times$ smaller index size than existing learned multi-dimensional indexes, in addition to up to 11$\times$ faster query performance and 170$\times$ smaller index size than optimally-tuned traditional indexes.
\end{abstract}

\maketitle


\section{Introduction}
\label{sec:intro}
Filtering through data is the foundation of any analytical database engine, and several advances over the past several years specifically target database filter performance.
For example, column stores~\cite{cstore} delay or entirely avoid accessing columns (i.e., dimensions) which are not relevant to a query, and they often sort the data by a single dimension in order to skip over records that do not match a query filter over that dimension. 

If data has to be filtered by more than one dimension, secondary indexes can be used. 
Unfortunately, their large storage overhead and the latency incurred by chasing pointers make them viable only when the predicate on the indexed dimension has a very high selectivity.
An alternative approach is to use (clustered) \emph{multi-dimensional} indexes; these may be tree-based data structures (e.g., k-d trees, R-trees, or octrees)
or a specialized sort order over multiple dimensions (e.g., a space-filling curve like Z-ordering or hand-picked hierarchical sort). 
Many state-of-the-art analytical database systems use multi-dimensional indexes or sort orders to improve the scan performance of queries with predicates over several columns~\cite{redshift,spark-sql,ibm-rtree}.

However, multi-dimensional indexes have significant drawbacks. 
First, these techniques are hard to tune and require an admin to carefully pick which dimensions to index, if any at all, and the order in which they are indexed. This decision must be revisited every time the data or workload changes, requiring extensive manual labor to maintain performance.
Second, there is no single approach (even if tuned correctly) that dominates all others~\cite{nathan2020flood}.

\begin{figure*}[]
    \includegraphics[width=\textwidth,clip]{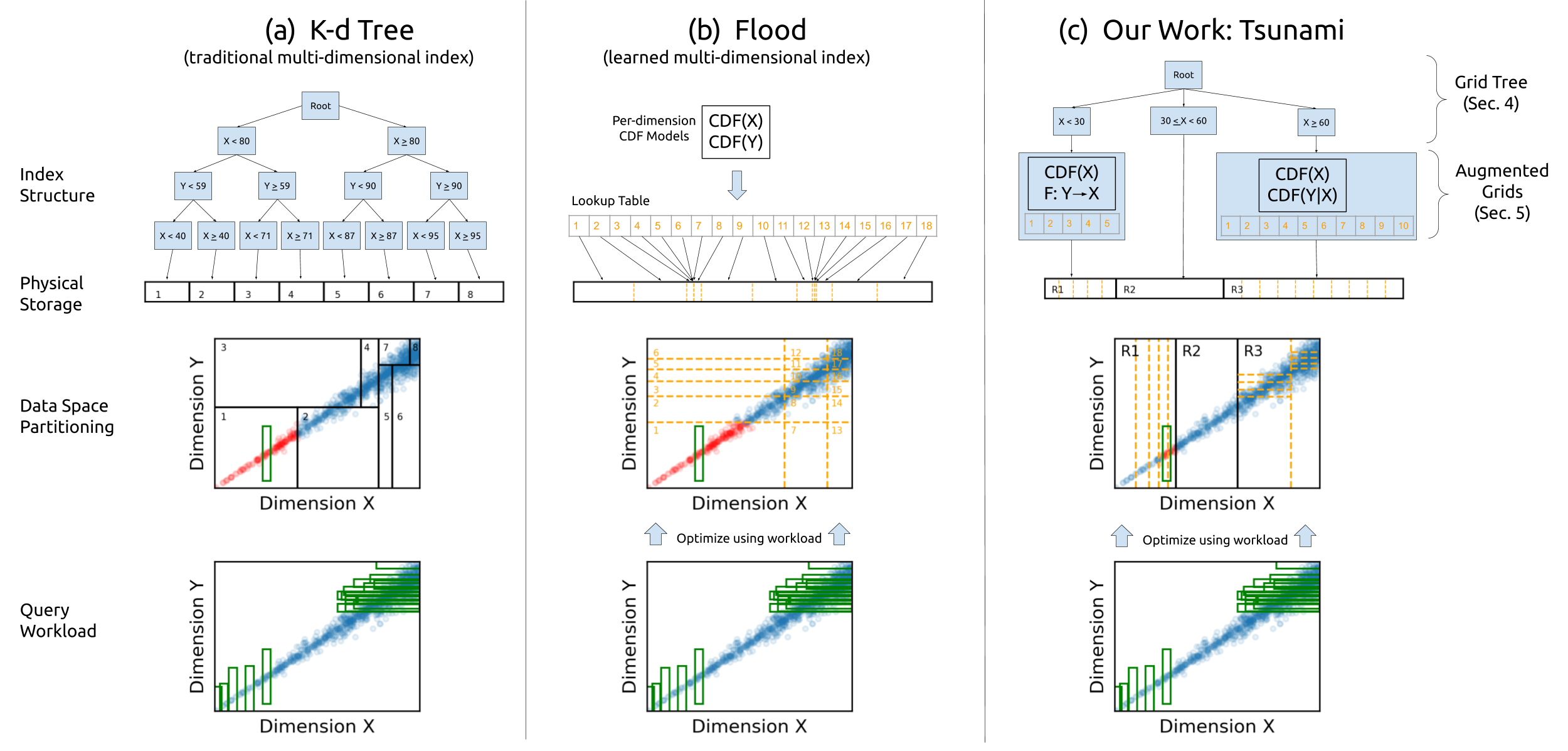}
    \caption{Indexes must identify the points that fall in the green query rectangle. To do so, they scan the points in red. (a) K-d tree guarantees equally-sized regions but is not optimized for the workload. (b) Flood is optimized using the workload but its structure is not expressive enough to handle query skew, and cells are unequally sized on correlated data. (c) \tsunami is optimized using the workload, is adaptive to query skew, and maintains equally-sized cells within each region.}
    \label{fig:design_overview}
\end{figure*}

To address the shortcomings of traditional indexes, recent work has proposed the idea of \textit{learned} multi-dimensional indexes~\cite{nathan2020flood,qdtree-sigmod,lisa-sigmod,zm-index,ml-index}.
In particular, Flood~\cite{nathan2020flood} is a  in-memory multi-dimensional index that automatically optimizes its structure to achieve high performance on a particular dataset and workload.
In contrast to traditional multi-dimensional indexes, such as the k-d tree, which are created entirely based on the data (see \Figure{design_overview}a), Flood divides each dimension into some number of partitions based on the observed data and workload (see \Figure{design_overview}b, explained in detail in \Section{background}). The Cartesian product of the partitions in each dimension form a grid. Furthermore, to reduce the index size, Flood uses models of the CDF of each dimension to locate the data.

However, Flood faces a number of limitations in real-world scenarios. 
First, Flood's grid cannot efficiently adapt to \textit{skewed} query workloads in which query frequencies and filter selectivities vary across the data space. Second, if dimensions are correlated, then Flood cannot maintain uniformly sized grid cells, which degrades performance and memory usage.


To address these limitations, we propose \tsunami, an in-memory read-optimized learned multi-dimensional index that extends the ideas of Flood with new data structures and optimization techniques.
First, \tsunami achieves high performance on skewed query workloads by using a lightweight decision tree, called a \gridtree, to partition space into non-overlapping regions in a way that reduces query skew.
Second, \tsunami achieves high performance on correlated datasets by indexing each region using an \augmentedgrid, which uses two techniques---\textit{functional mappings} and \textit{conditional CDFs}---to efficiently capture information about correlations. 

While recent work explored how correlation can be exploited to reduce the size of secondary indexes~\cite{hermit,correlationmaps}, our work goes much further. We demonstrate not only how to leverage correlation to achieve faster and more compact multi-dimensional indexes (in which the data is organized based on the index) rather than secondary indexes, but also how to integrate the optimization for query skew and data correlation into a full end-to-end solution.
\tsunami automatically optimizes the data storage organization as well as the multi-dimensional index structure based on the data and workload.

Like Flood~\cite{nathan2020flood}, Tsunami is a clustered in-memory read-optimized index over an in-memory column store. In-memory stores are increasingly popular due to lower RAM prices~\cite{cheap-ram} and our focus on reads reflects the current trend towards avoiding in-place updates in favor of incremental merges (e.g., RocksDB~\cite{rocksdb}). We envision that \tsunami could serve as the building block for a multi-dimensional in-memory key-value store or be integrated into commercial in-memory (offline) analytics accelerators like Oracle’s Database In-Memory (DBIM)~\cite{dbim}.

In summary, we make the following contributions:
\begin{CompactEnumerate}
\item We design and implement \tsunami, an in-memory read-optimized learned multi-dimensional index that self-optimizes to achieve high performance and robustness to correlated datasets and skewed workloads.
\item We introduce two data structures, the \gridtree and the \augmentedgrid, along with new optimization procedures that enable \tsunami to tailor its index structure and data organization strategy to handle data correlation and query skew.
\item We evaluate \tsunami against Flood, the original in-memory learned multi-dimensional index, as well as a number of traditional non-learned indexes, on a variety of workloads over real datasets. We show that \tsunami is up to 6$\times$ and 11$\times$ faster than Flood and the fastest optimally-tuned non-learned index, respectively. \tsunami is also adaptable to workload shift, and scales across data size, query selectivity, and dimensionality.
\end{CompactEnumerate}

In the remainder of this paper, we give background (\Section{background}), present an overview of \tsunami (\Section{tsunami}), introduce its two core components---\gridtree (\Section{gridtree}) and \augmentedgrid (\Section{augmentedgrid}), present experimental results
(\Section{eval}), review related work (\Section{related}), propose future work (\Section{future}), and conclude
(\Section{conclusion}).

\section{Background}
\label{sec:background}
\tsunami is an in-memory clustered multi-dimensional index for a single table. \tsunami aims to increase the throughput performance of analytics queries by decreasing the time needed to filter records based on range predicates. \tsunami supports queries such as:

\indent\indent\texttt{{\small SELECT SUM(R.X) \\ \indent\indent FROM MyTable \\ \indent\indent WHERE (a $\leq$ R.Y $\leq$ b) AND (c $\leq$ R.Z $\leq$ d)}}

\noindent where \texttt{SUM(R.X)} can be replaced by any aggregation. Records in a $d$-dimensional table can be represented as points in $d$-dimensional data space. For the rest of this paper, we use the terms \textit{record} and \textit{point} interchangeably. To place \tsunami in context, we first describe the k-d tree as an example of a traditional non-learned multi-dimensional index, and Flood, which originally proposed the idea of learned in-memory multi-dimensional indexing.

\subsection{K-d Tree: A Traditional Non-Learned Index}
\label{sec:kd_tree}
The k-d tree~\cite{kdtree} is a binary space-partitioning tree that recursively splits $d$-dimensional space based on the median value along each dimension, until the number of points in each leaf region falls below a threshold, called the page size. \Figure{design_overview}a shows a k-d tree over 2-dimensional data that has 8 leaf regions. The points within each region are stored contiguously in physical storage (e.g., a column store). By construction, the leaf regions have a roughly equal number of points.
To process a query (i.e., identify all points that match the query's filter predicates), the k-d tree traverses the tree to find all leaf regions that intersect the query's filter, then scans all points within those regions to identify points that match the filter predicates.

The k-d tree structure is constructed based on the data distribution but \textit{independently} of the query workload. That is, regardless of whether a region of the space is never queried or whether queries are more selective in some dimensions than others, the k-d tree would still build an index over all data points with the same page size and index overhead.
While other traditional multi-dimensional indexes split space in different ways~\cite{octree,rstar-tree,gridfile,amazon-zorder}, they all share the property that the index is constructed \textit{independent} of the query workload.

\subsection{Flood: A Learned Index}
\label{sec:flood}
In contrast, Flood~\cite{nathan2020flood} does optimize its layout based on the workload (see \Figure{design_overview}b). We first introduce how Flood works, then explain its two key advantages over traditional indexes, then discuss two key limitations it has.

Given a $d$-dimensional dataset, Flood first constructs compact models of the CDF of each dimension. The choice of modeling technique is orthogonal; Flood uses an RMI~\cite{rmi}, but one could also use a histogram or linear regression. Flood uses these models to divide the domain of each dimension into equally-sized \textit{partitions}: let $p_i$ be the number of partitions in each dimension $i\in[0,d)$. Then a point whose value in dimension $i$ is $x$ is placed into the $\lfloor CDF_i(x)\cdot p_i \rfloor$-th partition of dimension $i$, and similarly for all other dimensions. This guarantees that each partition in a given dimension has an equal number of points. When combined, the partitions of each dimension form a $d$-dimensional grid with $\prod_{i\in[0,d)}p_i$ \textit{cells}, which are ordered. The points within each cell are stored contiguously in physical storage.

Flood's query processing workflow has three steps, shown in \Figure{design_overview}b: (1) Using the per-dimension CDF models, identify the range of intersecting partitions in each dimension, and take the Cartesian product to identify the set of intersecting cells. (2) For each intersecting cell, identify the corresponding range in physical storage using a lookup table. (3) Scan all the points within those physical storage ranges, and identify the points that match all query filters.

\subsubsection{Flood's Strengths}
\label{sec:flood_strengths}
Flood has two key advantages over traditional indexes such as the k-d tree\footnote{Flood's minor third advantage, the \textit{sort dimension}, is orthogonal to our work.}. First, Flood can automatically tune its grid structure for a given query workload by adjusting the number of partitions in each dimension to maximize query performance. For example, in \Figure{design_overview}b, there are many queries in the upper-right region of the data space that have high selectivity over dimension Y. Therefore, Flood's optimization technique will place more partitions in dimension Y than dimension X, in order to reduce the number of points those queries need to scan. In other words, Flood learns which dimensions to prioritize over others and adjusts the number of partitions accordingly, whereas non-learned approaches do not take the workload into account and treat all dimensions equally.

Flood's second key advantage is its CDF models. The advantage of indexing using compact CDF models, as opposed to a tree-based structure such as a k-d tree, is lower overhead in both space and time: storing $d$ CDF models takes much less space than storing pointers and boundary keys for all internal tree nodes. It is also much faster to identify intersecting grid cells by invoking $d$ CDF models than by pointer chasing to traverse down a tree index.

The combination of these two key advantages allows Flood to outperform non-learned indexes by up to three orders of magnitude while using up to 50$\times$ smaller index size~\cite{nathan2020flood}, despite the simplicity of the grid structure.

\subsubsection{Flood's Limitations}
\label{sec:flood_limitations}
However, Flood has two key limitations. First, Flood only optimizes for the \textit{average} query, which results in degraded performance when queries are not uniform. For example, in \Figure{design_overview}b there are a few queries in the lower-left region of the data space that, unlike the many queries in the upper-right region, have high selectivity over dimension X. Since these queries are a small fraction of the total workload, Flood's optimization will not prioritize their performance.
As a result, Flood will need to scan a large number of points to create the query result (red points in \Figure{design_overview}b). Flood's uniform grid structure can only optimize for the average selectivity in each dimension and is not expressive enough to optimize for \textit{both} the upper-right queries and lower-left queries independently. The workload in \Figure{design_overview} is an example of a skewed workload. Query skew is common in real workloads: for example, queries often hit recent data more frequently than stale data, and operations monitoring systems only query for health metrics that are exceedingly low or high.

Second, Flood's model-based indexing technique can result in unequally-sized cells when data is correlated. In \Figure{design_overview}b, even though the CDF models guarantee that the three partitions over dimension X have an equal number of points, as do the six partitions over dimension Y, the 18 grid cells are unequally sized. This degrades performance and space usage (\Section{correlation_challenges}). Correlations are common in real data: for example, the price and distance of a taxi ride are correlated, as are the dates on which a package is shipped and received.

The goal of our work, \tsunami, is to maintain the two advantages of Flood---optimization based on the query workload and a compact/fast model-based index structure---while also addressing Flood's limitations in the presence of data correlations and query skew.

\section{\tsunami Design Overview}
\label{sec:tsunami}
\tsunami is a learned multi-dimensional index that is robust to data correlation and query skew. We first introduce the index structure and how it is used to process a query. We then provide an overview of the offline procedures we use to automatically optimize \tsunami's structure.

\NewPara{\tsunami Structure.}
\tsunami is a composition of two independent data structures: the \gridtree (\Section{gridtree}) and the \augmentedgrid (\Section{augmentedgrid}). The \gridtree is a space-partitioning decision tree that divides $d$-dimensional data space into some number of non-overlapping \textit{regions}. In \Figure{design_overview}c, the \gridtree divides  data space into three regions by splitting on dimension X.

Within each region, there is an \augmentedgrid. Each \augmentedgrid indexes the points that fall in its region. In \Figure{design_overview}c, Regions 1 and 3 each have their own \augmentedgrid. Region 2 is not given an \augmentedgrid because no queries intersect its region. An \augmentedgrid is essentially a generalization of Flood's index structure that uses additional techniques to capture correlations. In \Figure{design_overview}c, the \augmentedgrids use $F:Y\rightarrow X$ and $CDF(Y|X)$ instead of Flood's $CDF(Y)$ (explained in \Section{structures}).

\NewPara{\tsunami Query Workflow.}
\tsunami processes a query in three steps: (1) Traverse the \gridtree to find all regions that intersect the query's filter. (2) In each region, identify the set of intersecting \augmentedgrid cells (\Section{augmentedgrid}), then identify the corresponding range in physical storage using a lookup table. (3) Scan all the points within those physical storage ranges, and identify the points that match all query filters.

\NewPara{\tsunami Optimization.}
\tsunami's offline optimization procedure has two steps: (1) Optimize the \gridtree using the full dataset and sample query workload (\Section{gridtree_opt}). (2) In each region of the optimized \gridtree, construct an \augmentedgrid that is optimized over only the points and queries that intersect its region (\Section{augmentedgrid_opt}).

Intuitively, \tsunami separates the two concerns of query skew and data correlations into its two component structures, \gridtree and \augmentedgrid, respectively. Each structure is optimized in a way that addresses its corresponding concern. We now describe each structure in detail.
\section{\gridtree}
\label{sec:gridtree}
In this section, we first discuss the performance challenges posed by skewed workloads. We then formally define query skew, and we describe \tsunami's solution for mitigating query skew: the \gridtree.

\subsection{Challenges of Query Skew}
\label{sec:query_skew_challenges}
\begin{figure}[]
    \subfloat{
        \includegraphics[width=0.48\columnwidth,clip]{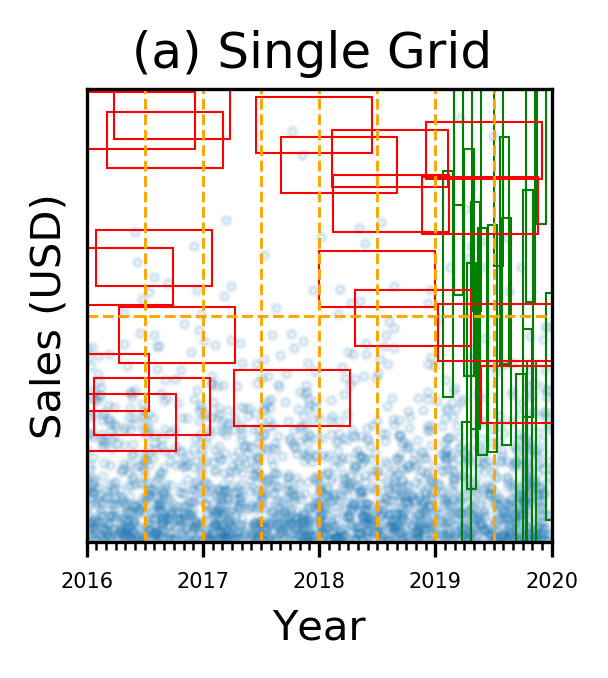}
        \label{fig:skewed_workload_a}
    }
    ~
    \subfloat{
        \includegraphics[width=0.48\columnwidth,clip]{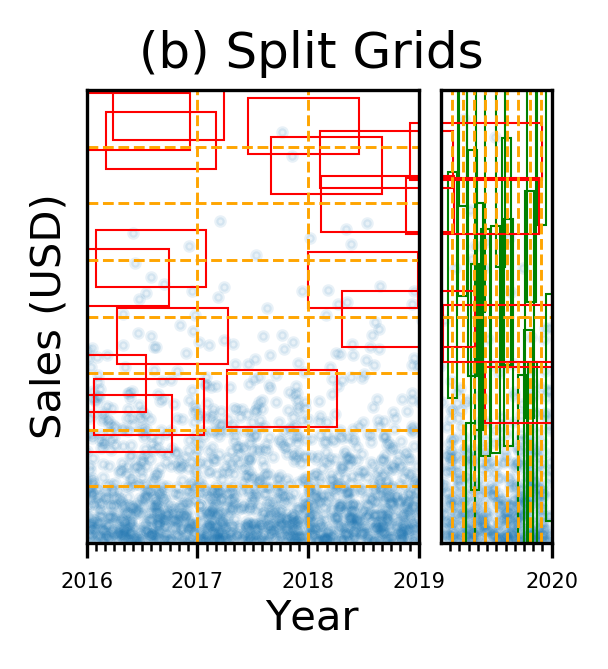}
        \label{fig:skewed_workload_b}
    }
    \caption{A single grid cannot efficiently index a skewed query workload, but a combination of non-overlapping grids can. We use this workload as a running example.}
    \label{fig:skewed_workload}
\end{figure}

A query workload is skewed if the characteristics of queries (e.g., frequency or selectivity) vary in different parts of the data space. \Figure{skewed_workload_a} shows an example of sales data from 2016 to 2020. Points are uniformly distributed in time. The query workload is composed of two distinct query ``types'': the red queries $Q_r$ filter uniformly over one-year spans, whereas the green queries $Q_g$ filter over one-month spans only over the last year. If we were to impose a grid over the data space, we intuitively would want many partitions over the past year in order to obtain finer granularity for $Q_g$, whereas partitions prior to 2019 should be more widely spaced, because $Q_r$ does not require much granularity in time. However, with a single grid it is not possible accommodate both while maintaining an equal number of points in each partition (\Figure{skewed_workload_a}).

Instead, we can split the data space into two regions: before 2019 and after 2019 (\Figure{skewed_workload_b}). Each region has its own grid, and the two grids are independent. The right region can therefore tailor its grid for $Q_g$ by creating many partitions over time. On the other hand, the left region does not need to worry about $Q_g$ at all and places few partitions over time, and can instead add more partitions over the sales dimension. This intuition drives our solution for tackling query skew.

\subsection{Reducing Query Skew with a \gridtree}
We first formally define query skew. We then describe at a high level how \gridtree tackles query skew and how to process queries using the \gridtree. We then describe how to find the optimal \gridtree for a given dataset and query workload. We use the terminology in \Table{grid_tree_terms}.

\begin{table}[]
\centering
\caption{Terms used to describe the \gridtree}
\vspace{-1em}
\small
\label{tab:grid_tree_terms}
\begin{tabular}{@{}lll@{}}
\toprule
    \textbf{Term}     & \textbf{Description}   \\
\midrule
    $d$ & Dimensionality of the dataset \\
    $S$ & $d$-dimensional data space: $[0, X_0) \times \cdots \times [0, X_{d-1})$ \\
    $Q$ & Set of queries \\
    $Uni_i(a,b)$ & Uniform distribution over $[a,b)$ in dimension $i\in[0,d)$ \\
    $PDF_i(Q,a,b)$ & Empirical PDF of queries $Q$ over $[a,b)$ in dimension $i$ \\
    $Hist_i(Q,a,b,n)$ & Approximate PDF of queries $Q$ over range $[a,b)$ \\
    & in dimension $i$ using a histogram with $n$ bins \\
    $EMD(P_1,P_2)$ & Earth Mover's Distance between distributions $P_1,P_2$ \\
    $Skew_i(Q,a,b)$ & Skew of query set $Q$ over range $[a,b)$ in dimension $i$ \\
\bottomrule
\end{tabular}
\end{table}

\begin{figure}[]
    \centering
    \includegraphics[width=0.8\columnwidth]{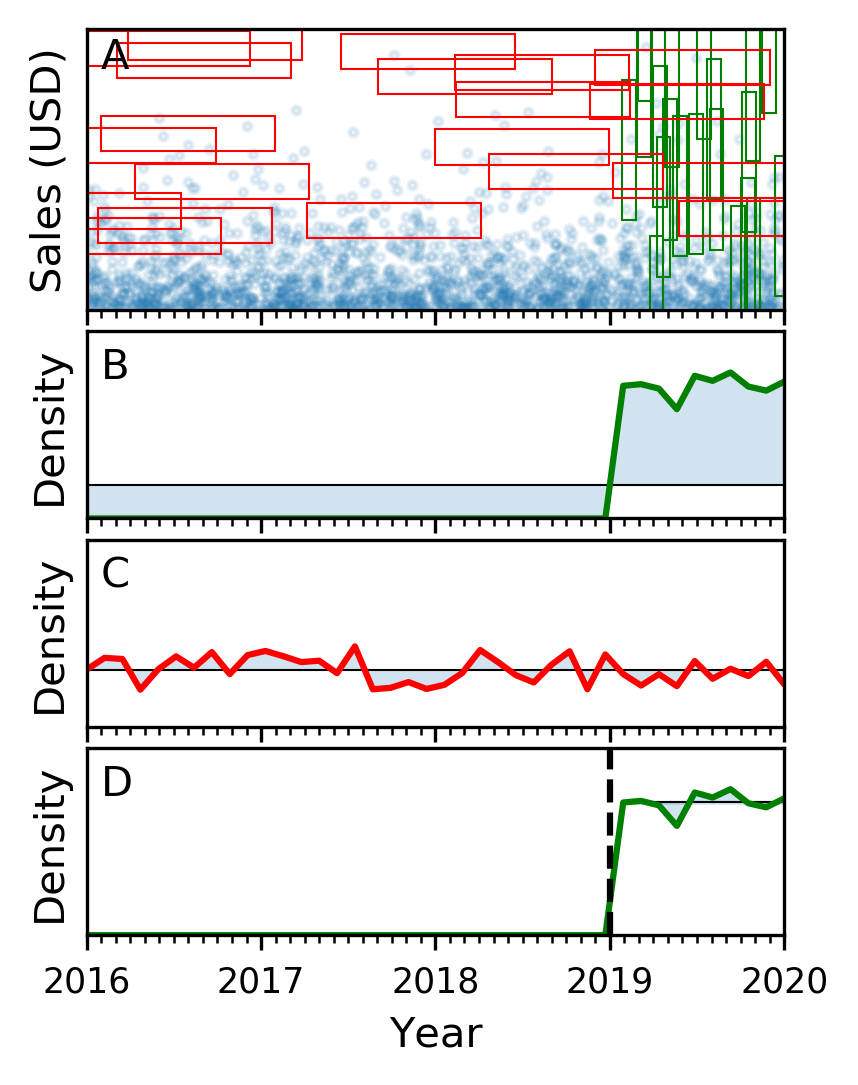}
    \caption{Query skew is computed independently for each query type ($Q_g$ and $Q_r$) and is defined as the statistical distance between the empirical PDF of the queries and the uniform distribution.}
    \label{fig:query_skew}
\end{figure}

\subsubsection{Definition of Query Skew}
The skew of a set of queries $Q$ with respect to a range $[a,b)$ in dimension $i$ is
\begin{align*}
    Skew_i(Q,a,b) = EMD\left(Uni_i(a,b), PDF_i(Q,a,b)\right)
\end{align*}
where $Uni_i(a,b)$ is a uniform distribution over $[a,b)$ and $PDF_i(Q,a,b)$ is the empirical PDF of queries in $Q$ over $[a,b)$. Each query contributes a unit mass to the PDF, spread over its filter range in dimension $i$. $EMD$ is the Earth Mover's Distance, which is a measure of the distance between two probability distributions.

\Figure{query_skew}a shows the same data and workload as in \Figure{skewed_workload}. \Figure{query_skew}b-c show the PDF of $Q_g$ and $Q_r$, respectively. The skew is intuitively visualized (though not technically equal to) the shaded area between the PDF and the uniform distribution. Although $Q_g$ is highly skewed over the time dimension, \Figure{query_skew}d shows that by splitting the time domain at 2019, we can reduce the skew of $Q_g$ because $Skew_{Year}(Q_g,2016,2019)$ and $Skew_{Year}(Q_g,2019,2020)$ are low.

In concept, $PDF_i(Q,a,b)$ is a continuous probability distribution. However, in practice we approximate $PDF_i(Q,a,b)$ using a histogram: we discretize the range $[a,b)$ into $n$ bins. If a query $q$'s filter range intersects with $m$ contiguous bins, then it contributes $1/m$ mass to each of the bins. Therefore, the total histogram mass will be $|Q|$. We call this histogram $Hist_i(Q,a,b,n)$.

In this context, a probability distribution over a range of histogram bins $[x,y)$, where $0\le x< y\le n$, is a $(y-x)$-dimensional vector. We can concretely compute skew over the bins $[x,y)$:
\begin{align*}
    Uni_i(Q,x,y)[j] &= \frac{\sum_{x\le k<y}Hist_i(Q,a,b,n)[k]}{y-x} \qquad \text{for $x\le j < y$} \\
    PDF_i(Q,x,y)[j] &= Hist_i(Q,a,b,n)[j] \qquad \text{for $x\le j < y$} \\
    Skew_i(Q,x,y) &= EMD\left(Uni_i(Q,x,y), PDF_i(Q,x,y)\right)
\end{align*}
We store the bin boundaries of the histogram, so there is a simple mapping function from a value $a$ to its bin $x$. Therefore, throughout this section, we will use $Skew_i(Q,a,b)$ and $Skew_i(Q,x,y)$ interchangeably.

\subsubsection{\gridtree Design}
Given a query workload that is skewed over a data space, the aim of the \gridtree is to divide the data space into a number of non-overlapping \textit{regions} so that within each region, there is little query skew.

The \gridtree is a space-partitioning decision tree, similar to a k-d tree. Each internal node of the \gridtree divides space based on the values in a particular dimension, called the split dimension $d_s$. Unlike a k-d tree, which is a binary tree, internal nodes of the \gridtree can split on more than one value. If an internal node splits on values $V = \{v_1,\ldots,v_k\}$, then the node has $k+1$ children. To process a query, we traverse the \gridtree to find all regions that intersect with the query's filter predicates. If there is an index over the points in that region (e.g., an \augmentedgrid), then we delegate the query to that index and aggregate the returned results. If there is no index for the region, we simply scan all points in the region.

Note that the \gridtree is not meant to be an end-to-end index. Instead, the \gridtree's purpose is to efficiently reduce query skew, while using low memory. This way, the user is free to use any indexing scheme within each region, without worrying about intra-region query skew.

\subsection{Optimizing the \gridtree}
\label{sec:gridtree_opt}
Given a dataset and sample query workload, our optimization goal is to reduce query skew as much as possible while maintaining a small and lightweight \gridtree.
We present the high-level optimization algorithm, then dive into details. Our procedure is as follows: (1) Group queries in the sample workload into some number of clusters, which we call query \textit{types} (\Section{clustering}). (2) Build the \gridtree in a greedy fashion. Start with a root node that is responsible for the entire data space $S$. Recursively, for each node $N$ responsible for data space $S_N$, pick the split dimension $d_s \in [0,d)$ and the set of split values $V = \{v_1,\ldots,v_k\}$ that most reduce query skew (\Section{split_selection}). $d_s$ and $V$ define $k+1$ non-overlapping sub-spaces of $S_N$. Assign a child node to each of the $k+1$ sub-spaces and recurse for each child node. If a node $N$ has low query skew (\Section{split_selection}), or has below a minimum threshold number of intersecting points or queries, then it stops recursing and becomes a leaf node, representing a \textit{region}.

\subsubsection{Clustering Query Types}
\label{sec:clustering}
It is not enough to consider the query skew of the entire query set $Q$ as a whole, because queries within this set have different characteristics and therefore are best indexed in different ways. For example, we showed in \Figure{skewed_workload} that $Q_g$ and $Q_r$ are best indexed with different partitioning schemes. Considering all queries as a whole can mask the effects of skew because the skews of different query types can cancel each other out.

Therefore, we cluster queries into \textit{types} that have similar selectivity characteristics. First, queries that filter over different sets of dimensions are automatically placed in different types. For each group of queries that filter over the same set of $d'$ dimensions, we transform each query into a $d'$-dimensional embedding in which each value is set to the filter selectivity of the query over a particular dimension. We run DBSCAN over the $d'$-dimensional embeddings with eps set to 0.2 (this worked well for all our experiments and we never tuned it). DBSCAN automatically determines the number of clusters. The choice of clustering algorithm is orthogonal to the \gridtree design.

Real query workloads have patterns and can usually be divided into types. For example, many analytic workloads are composed of query templates, for which the dimensions filtered and rough selectivity remains constant, but the specific predicate values vary. However, even if there are no patterns in the workload, the \gridtree is still useful because there can still be query skew over a single query type (i.e., query frequency varies in different parts of data space).

From now on, we assume that if the query set $Q$ is composed of $t$ query types, then we can divide $Q$ into $t$ subsets $Q_1,\ldots,Q_t$. For example, in \Figure{query_skew} there are 2 types, $Q_r$ and $Q_g$. Note that each query can only belong to one query type, but queries in different types are allowed to overlap in data space. We now redefine skew:
\begin{align*}
    Skew_i(Q,a,b) = \sum_{1\le i \le t}Skew_i(Q_t,a,b)
\end{align*}

\subsubsection{Selecting the Split Dimension and Values}
\label{sec:split_selection}
Given a \gridtree node $N$ over a data space $S_N$ and a set of queries $Q$ that intersects with $S_N$, our goal is to find the split dimension $d_s$ and split values over that dimension $V = \{v_1,\ldots,v_k\}$ that achieve the largest reduction in query skew. For a dimension $i\in[0,d)$ and split values $V$, the reduction in query skew is defined as
\begin{align*}
    R_i(Q,0,X_d,V) &= Skew_i(Q,0,X_d) - \Big[Skew_i(Q,0,v_1)  \\ &\quad + Skew_i(Q,v_k,X_d) + \sum_{1\le i< k} Skew_i(Q,v_i,v_{i+1}) \Big]
\end{align*}
Note that skew reduction is defined \textit{per dimension}, not over all $d$ dimensions simultaneously. Therefore, we independently find the largest skew reduction $Rmax_i=\max_{V}(R_i)$ for each dimension $i\in[0,d)$ (explained next), then pick the split dimension $d_s=\argmax_{i}(Rmax_i)$.

For example, in \Figure{query_skew}, $Skew_{Sales}$ is already low because both query types are distributed relatively uniformly over Sales, so $Rmax_\text{Sales}$ is low. On the other hand, $Skew_{Year}$ is high. We can achieve very large $Rmax_\text{Year}$ using $V=\{2019\}$. Therefore, we select $d_s=\text{Year}$ and $V=\{2019\}$.

If $\max_{i}(Rmax_i)$ is below some minimum threshold (by default 5\% of $|Q|$) or if $S_N$ intersects below a minimum threshold of points or queries (by default 1\% of the total points or queries in the entire data space), then $d_s$ is rejected and $N$ becomes a leaf \gridtree node.

\begin{figure}[]
    \centering
    \includegraphics[width=0.8\columnwidth]{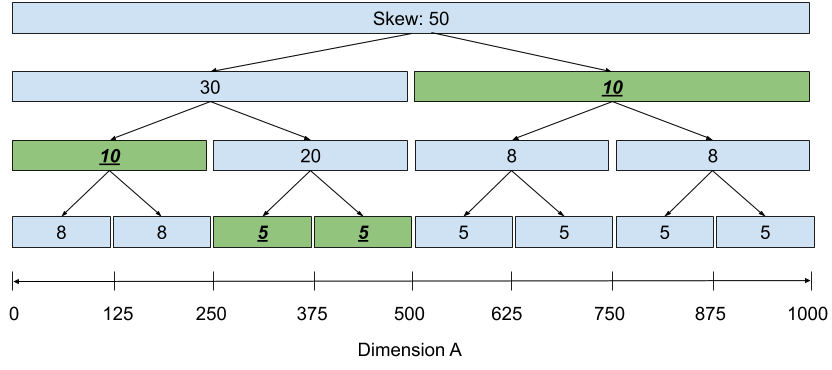}
    \caption{Skew tree over the range $[0, 1000)$ with eight leaf nodes. The covering set that achieves lowest combined skew is shaded green. Based on the boundaries of the covering set, we extract the split values $V=\{250, 375, 500\}$.}
    \label{fig:skew_tree}
\end{figure}
We now explain how to find the split values $V$ that maximize $R_{d_s}$ for each candidate split dimension $d_s \in [0, d)$. We introduce a data structure called the \textit{skew tree}, which is simply a tool to help find the optimal $V$; it is never used when running queries. The skew tree is a balanced binary tree (\Figure{skew_tree}). Each node represents a range over the domain of dimension $d_s$. The root node represents the entire range $[0, X_{d_s})$, and every node represents the combined ranges of the nodes in its subtree. A skew tree node whose range is $[a, b)$ will store the value $Skew_{d_s}(Q,a,b)$. In other words, each skew tree node stores the query skew over the range it represents.

Creating the skew tree requires $Hist_{d_s}(Q,0,X_{d_s})$. By default, we instantiate the histogram with 128 bins. Note that we are unable to compute a meaningful skew over a single histogram bin: $Skew_{d_s}(Q,x,x+1)$ is always zero, because a single bin has no way to differentiate the uniform distribution from the query PDF. Therefore, the skew tree will only have 64 leaf nodes. However, if there are fewer than 128 unique values in dimension $d_s$, we create a bin for each unique value. In this case, there is truly no skew within each histogram bin, so the skew tree has as many leaf nodes as unique values in $d_s$, and the skew at each leaf node is 0.

A set of skew tree nodes is called \textit{covering} if their represented ranges do not intersect and the union of their represented ranges is $[0,X_{d_s})$. We want to solve for the covering set with minimum combined query skew. This is simple to do via dynamic programming in two passes over the skew tree nodes: in the first pass, we start from the leaf nodes and work towards the root node, and at each node we annotate the minimum combined query skew achievable over the node's subtree. In the second pass, we start from the root and work towards the leaves, and check if a node's skew is equal to the annotated skew: if so, the node is part of the optimal covering set. The boundaries between the ranges of nodes in the optimal covering set form $V$.

As a final step, we do a single ordered pass over all the nodes in the covering set, in order of the range they represent, and merge nodes if the query skew of the combined node is not more than a constant factor (by default, 10\%) larger than the sum of the individual query skews. For example, in \Figure{skew_tree} if $Skew_A(Q,0,375)<15\cdot1.1$, then the first two nodes of the covering set would be merged, and $250$ would be removed as a split value. This step counteracts the fact that the binary tree may split at superfluous points, and it also acts as a regularizer that prevents too many splits.
\section{\augmentedgrid}
\label{sec:augmentedgrid}

In this section, we describe the challenges posed by data correlations, and we introduce our solution to address those challenges: the \augmentedgrid.
Note that the \gridtree (\Section{gridtree}) optimizes only for query skew reduction, and the points within each \textit{region} might still display correlation.

\subsection{Challenges of Data Correlation}
\label{sec:correlation_challenges}
We broadly define a pair of dimensions $X$ and $Y$ to be correlated if they are not independent, i.e., if $CDF(X) \ne CDF(X|Y)$ and vice versa.
In the presence of correlated dimensions, it is not possible to impose a grid that has equally-sized cells by partitioning each dimension independently (see \Figure{design_overview}b). As a result, points will be clustered into a relatively few number of  cells, so any query that hits one of those cells will likely scan many more points than necessary.

One way to mitigate this issue is by increasing the number of partitions in each dimension, to form more fine-grained cells. 
However, increasing the number of cells would counteract the two advantages of grids over trees: (1) Space overhead increases rapidly (e.g., doubling the number of partitions in each dimension increases index size by $2^d$). (2) Time overhead also increases, because each cell incurs a lookup table lookup. Therefore, simply making finer-grained grids is not a scalable solution to data correlations.

\subsection{A Correlation-Aware Grid}
\label{sec:structures}
\tsunami handles data correlations while maintaining the time and space advantage of grids by augmenting the basic grid structure with new partitioning strategies that allow it to partition dimensions \textit{dependently} instead of independently. We first provide a high level description of the \augmentedgrid, then dive into details.

An \augmentedgrid is a grid in which each dimension $X \in [0, d)$ is divided into $p_X$ partitions and uses one of three possible strategies for creating its partitions: (1) We can partition $X$ independently of other dimensions, uniformly in $CDF(X)$. This is what Flood does for every dimension. (2) We can remove $X$ from the grid and transform query filters over $X$ into filters over some other dimension $Y\in [0, d)$ using a functional mapping $F:X\rightarrow Y$ (\Section{functional_mapping}). (3) We can partition $X$ dependent on another  dimension $Y\in [0, d)$, uniformly in $CDF(X|Y)$ (\Section{ccdf}).

\begin{table}[]
\small
\begin{tabular}{llll}
\toprule
     \textbf{Ex. skeleton} & \multicolumn{3}{c}{$[X, Y|X, Z]$ (i.e., $CDF(X)$, $CDF(Y|X)$, and $CDF(Z)$)}   \\
\midrule
\textbf{One hop away} & $[X, Y, Z]$    & $[X, Y|Z, Z]$    & $[X, Y\rightarrow X, Z]$  \\
& $[X, Y\rightarrow Z, Z]$ & $[X, Y|X, Z|X]$    & $[X, Y|X, Z\rightarrow X]$   \\
\bottomrule
\end{tabular}
\centering
\caption{Example skeleton over dimensions $X,Y,Z$, and all skeletons one ``hop'' away. Restrictions are explained in \Section{functional_mapping} and \Section{ccdf} (e.g., $[X\rightarrow Z, Y|X, Z]$ is not allowed).}
\label{tab:skeletons}
\vspace{-1em}
\end{table}

A specific instantiation of partitioning strategies for all dimensions is called a \textit{skeleton}. \Table{skeletons} shows an example. We ``flesh out'' the skeleton by setting the number of partitions in each dimension to create a concrete instantiation of an \augmentedgrid. Therefore, an \augmentedgrid is uniquely defined by the combination of its skeleton $S$ and number of partitions in each dimension $P$.

\subsubsection{Functional Mappings}
\label{sec:functional_mapping}
A pair of dimensions $X$ and $Y$ is monotonically correlated if as values in $X$ increase, values in $Y$ only move in one direction. Linear correlations are one subclass of monotonic correlations. For monotonically correlated $X$ and $Y$, we conceptually define a mapping function as a function $F : \mathbb{R}^2 \rightarrow \mathbb{R}^2$ that takes a range $[Y_{min},Y_{max}]$ over dimension $Y$ and maps it to a range $[X_{min},X_{max}]$ over dimension $X$ with the guarantee that any point whose value in dimension $Y$ is in $[Y_{min},Y_{max}]$ will have a value in dimension $X$ in $[X_{min},X_{max}]$. In this case, we call $Y$ the \textit{mapped dimension} and we call $X$ the \textit{target dimension}. For simplicity, we place a restriction: a target dimension cannot itself be a mapped dimension. Similar ideas were proposed in~\cite{correlationmaps,hermit}.

Concretely, we implement the mapping function as a simple linear regression $LR$ trained to predict $X$ from $Y$, with lower and upper error bounds $e_l$ and $e_u$. Therefore, a functional mapping is encoded in four floating point numbers and has negligible storage overhead. Given a range $[Y_{min},Y_{max}]$, the mapping function produces $X_{min} = Y_{min}-e_l$ and $X_{max} = Y_{max}+e_u$. Note that the idea of functional mappings can generalize to all monotonic correlations, as in~\cite{hermit}. However, in our experience the vast majority of monotonic correlations in real data are linear, so we use linear regressions for simplicity.

\begin{figure}[]
    \centering
    \includegraphics[width=0.8\columnwidth]{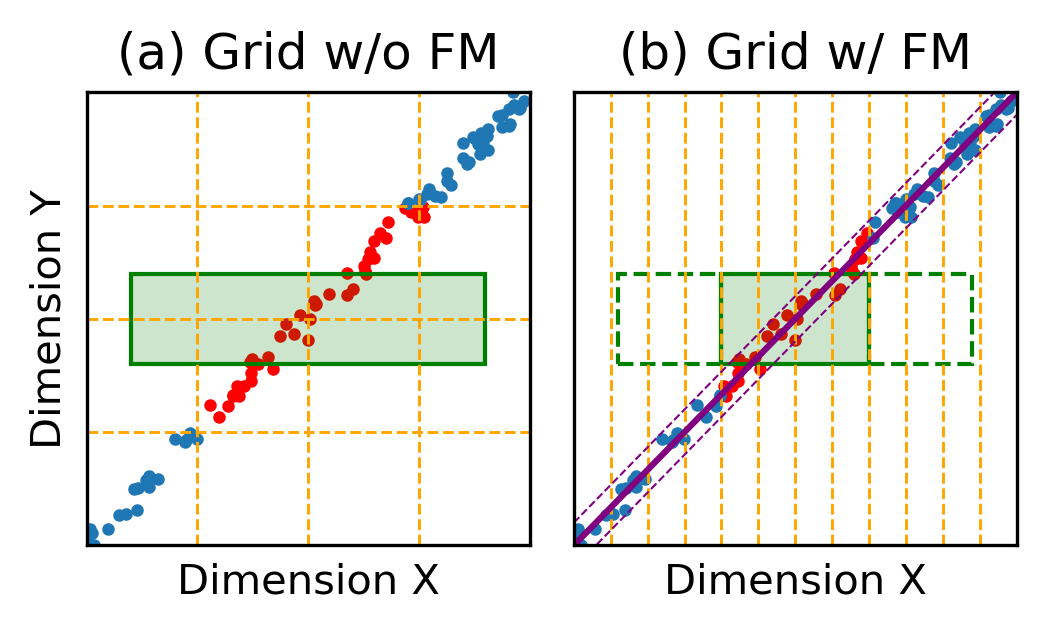}
    \caption{Functional mapping creates equally-sized cells and reduces scanned points for tight monotonic correlations. The query is in green, scanned points are red, and the mapping function is purple, with error bounds drawn as dashed lines.}
    \label{fig:functional_mapping}
    \vspace{-1em}
\end{figure}

Given a functional mapping, any range filter predicate $(y_0 \leq Y \leq y_1)$ over dimension $Y$ can be transformed into a semantically equivalent predicate $(x_0 \leq X \leq x_1)$ over dimension $X$, where $(x_0, x_1) = F(y_0, y_1)$. This gives us the opportunity to completely remove the mapped dimension from the $d$-dimensional grid, to obtain equally-sized cells.
\Figure{functional_mapping} demonstrates the benefits of functional mapping. The grid without functional mapping has unequally-sized cells, which results in many points scanned. On the other hand, the grid with functional mapping has equally-sized cells and is furthermore able to ``shrink'' the size of the query to a semantically equivalent query by \textit{inducing} a narrower filter over dimension X using the mapping function. This results in fewer points scanned.

\subsubsection{Conditional CDFs}
\label{sec:ccdf}
Functional mappings are only useful for tight monotonic correlations. Otherwise, the error bounds would be too large for the mapping to be useful. For loose monotonic correlations or generic correlations, we instead use conditional CDFs. For a pair of generically correlated dimensions $X$ and $Y$, we partition $X$ uniformly in $CDF(X)$ and we partition $Y$ uniformly in $CDF(Y|X)$, resulting in equally-sized cells. In this case, we call $X$ the \textit{base dimension} and $Y$ the \textit{dependent dimension}. For simplicity, we place restrictions: a base dimension cannot itself be a mapped dimension or a dependent dimension.

Concretely, if there are $p_X$ and $p_Y$ partitions over $X$ and $Y$ respectively, we implement $CDF(Y|X)$ by storing $p_X$ histograms over $Y$, one for each partition in $X$. When a query filters over $Y$, we first find all intersecting partitions in $X$, then for each $X$ partition independently invoke $CDF(Y|X)$ to find the intersecting partitions in $Y$. The storage overhead is proportional to $p_Xp_Y$, which is minimal compared to the existing overhead of the grid's lookup table, which is proportional to $\prod_{i\in[0,d)}p_i$.

\begin{figure}[]
    \centering
    \subfloat{
        \includegraphics[width=0.8\columnwidth,clip]{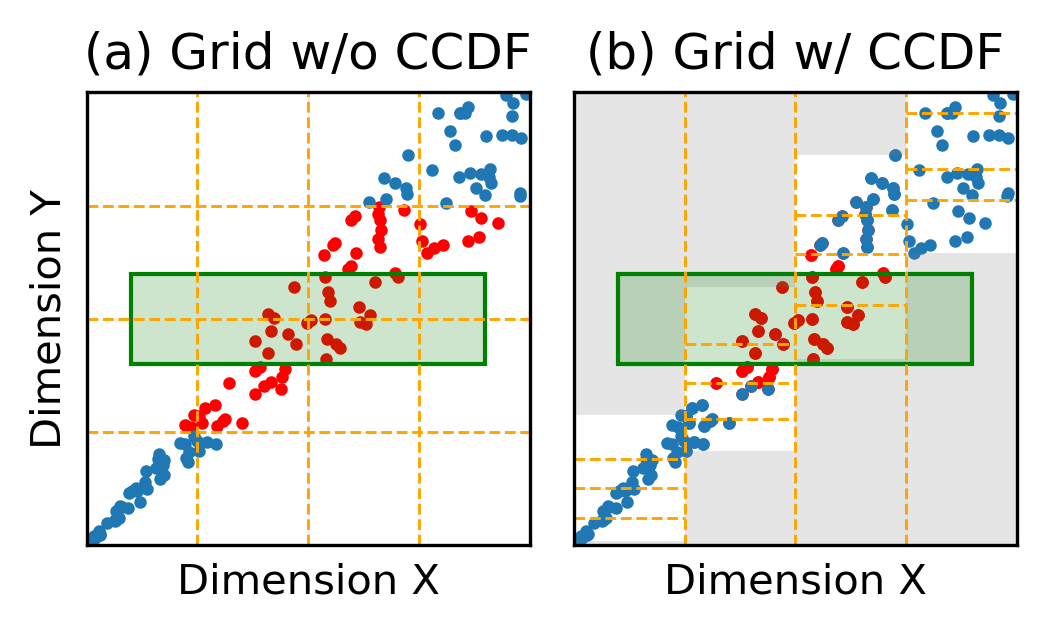}
    }
    \caption{Conditional CDFs create equally-sized cells and reduce scanned points for generic correlations. The query is in green, and scanned points are in red.}
    \label{fig:nonuniform_grid}
    \vspace{-1em}
\end{figure}

\Figure{nonuniform_grid} shows an example of using conditional CDFs. Both grids have $p_X=p_Y=4$. By partitioning $Y$ using $CDF(Y|X)$, the grid on the right has staggered partition boundaries, which create equally-sized cells and results in fewer points scanned. Additionally, the regions outside the cells (shaded in gray) are guaranteed to have no points, which allows the query to avoid scanning the first and last partitions of $X$, even though they intersect the query.

\subsection{Optimizing the \augmentedgrid}
\label{sec:augmentedgrid_opt}
Given a dataset and sample query workload, our optimization goal is to find the best \augmentedgrid, i.e., the settings of the parameters $(S, P)$ that achieves lowest average query time over the sample workload, where $S$ is the skeleton and $P$ is the number of partitions in each dimension.

This optimization problem is challenging in two ways: (1) For a specific setting of $(S,P)$, we cannot know the average query time without actually running the queries, which can be very time-intensive. Therefore, we create a cost model to predict average query time, and we optimize for lowest average \textit{predicted} query time (\Section{cost_model}). (2) The search space over skeletons is exponentially large. For each dimension, there are $O(d)$ possible partitioning strategies, since there are up to $d-1$ choices for the other dimension in a functional mapping or conditional CDF. Therefore, the search space of skeletons has size $O(d^d)$. To efficiently navigate the joint search space of $(S,P)$, we use \textit{adaptive gradient descent} (\Section{agd}).

\subsubsection{Cost Model}
\label{sec:cost_model}
We use a simple analytic linear cost model to predict the runtime of a query $q$ on dataset $D$ and an instantiation of the \augmentedgrid with parameters $(S,P)$:
\begin{align*}
    \text{Time} &= w_0 (\text{\# cell ranges}) + w_1 (\text{\# scanned points})(\text{\# filtered dims})
\end{align*}
We now explain each term of this model. A set of adjacent cells in physical storage is called a \textit{cell range}. Instead of doing a lookup on the lookup table for every intersecting cell, we only look up the first and last cell of a cell range. Furthermore, skipping to each new cell range in physical storage likely incurs a cache miss. $w_0$ represents the time to do a lookup and the cache miss of accessing the range in physical storage.

The $w_1$ term models the time to scan points (e.g., all red points in previous figures). Since data is stored in a column store, only the dimensions filtered by the query need to be accessed. $w_1$ represents the time to scan a single dimension of a single point.

Importantly, the features of this cost model can be efficiently computed or estimated: the number of cell ranges is easily computed from $q$ and $(S,P)$. The number of filtered dimensions is obvious from $q$. The number of scanned points is estimated using $q$, $(S,P)$, and a sample of $D$.

Note that we do not model the time to actually perform the aggregation after finding the points that intersect the query rectangle. This is because aggregation is a fixed cost that must be incurred regardless of index choice, so we ignore it when optimizing.

\subsubsection{Adaptive Gradient Descent}
\label{sec:agd}
We find the $(S,P)$ that minimizes average query time, as predicted by the cost model, using adaptive gradient descent (AGD). We first enumerate AGD's high level steps, then provide details for each step. AGD is an iterative algorithm that jointly optimizes $S$ and $P$:
\begin{CompactEnumerate}
\item Using heuristics, initialize $(S_0,P_0)$.
\item From $(S_0,P_0)$, take a gradient descent step over $P_0$ using the cost model as the objective function, which gives us $(S_0,P_1)$. 
\item From $(S_0,P_1)$, perform a local search over skeletons to find the skeleton $S'$ that minimizes query time for $(S',P_1)$. Set $S_1=S'$. It may be that $S'=S_0$, that is, the skeleton does not change in this step.
\item Repeat steps 2 and 3 starting from $(S_1,P_1)$ until we reach a minimum average query time.
\end{CompactEnumerate}

In step 1, we first initialize $S$, then $P$. We make a best guess at the optimal skeleton using heuristics: for each dimension $X$, use a functional mapping to dimension $Y$ if the error bound is below 10\% of $Y$'s domain. Else, partition using $CDF(X|Y)$ if not doing so would result in more than 25\% of cells in the $XY$ grid hyperplane being empty. Else, partition $X$ independently using $CDF(X)$. Given the initial $S$, we initialize $P$ proportionally to the average query filter selectivity in each grid dimension (i.e., excluding mapped dimensions).

In step 2, we take advantage of the insight that the cost model is relatively smooth in $P$: changing the number of partitions in a dimension usually smoothly increases or decreases the cost. Therefore, we take the numerical gradient over $P$ at $(S,P)$ and take a step in the gradient direction.

In step 3, we take advantage of the insight that an incremental change in $P$ is unlikely to cause the skeleton $S'$ to differ greatly from $S$. Therefore, step 3 will only search over $S'$ that can be created by changing the partitioning strategy for a single dimension in $S$ (e.g., skeletons one ``hop'' away in \Table{skeletons}).

While we could conceivably use black box optimization methods such as simulated annealing to optimize $(S,P)$, AGD takes advantage of the aforementioned insights into the behavior of the optimization and is therefore able to find lower-cost \augmentedgrids, which we confirm in \Section{components_results}.

\section{Evaluation}
\label{sec:eval}
We first describe the experimental setup and then present the results of an
in-depth experimental study that compares \tsunami with Flood and several other indexing methods on a variety of datasets and workloads. Overall, this evaluation shows that:
\begin{CompactEnumerate}
	\item \tsunami is consistently the fastest index across tested datasets and workloads. It achieves up to 6$\times$ higher query throughput than Flood and up to 11$\times$ higher query throughput than the fastest optimally-tuned non-learned index. Furthermore, \tsunami has up to 8$\times$ smaller index size than Flood and up to 170$\times$ smaller index size than the fastest non-learned index (\Section{overall_results}).
	\item \tsunami can optimize its index layout and reorganize the records quickly for a new query distribution, typically in under 4 minutes for a 300 million record dataset (\Section{adaptibility_results}).
	\item \tsunami's performance advantage over other indexes scales with dataset size, selectivity, and dimensionality (\Section{scalability_results}).
\end{CompactEnumerate}

\subsection{Implementation and Setup}
We implement \tsunami in C++ and perform optimization in Python. We perform our query performance evaluation via single-threaded experiments on an Ubuntu Linux machine with Intel Core i9-9900K 3.6GHz CPU and 64GB RAM. Optimization and data sorting for index creation are performed in parallel for \tsunami and all baselines.

All experiments use 64-bit integer-valued attributes. Any string values are dictionary encoded prior to evaluation. Floating point values are typically limited to a fixed number of decimal points (e.g., 2 for price values). We scale all values by the smallest power of 10 that converts them to integers.

Evaluation is performed on data stored in a custom column store with one scan-time optimization: if the range of data being scanned is \emph{exact}, i.e., we are guaranteed ahead of time that all elements within the range match the query filter, we skip checking each value against the query filter. For common aggregations, e.g. \texttt{\small COUNT}, this removes unnecessary accesses to the underlying data.

We compare \tsunami to other solutions implemented on the same column store, with the same optimizations, if applicable:
\begin{CompactEnumerate}
    \item \emph{Clustered Single-Dimensional Index}: Points are sorted by the most selective dimension in the query workload. If a query filter contains this dimension, we locate the endpoints using binary search. Otherwise, we perform a full scan.
    \item The \emph{Z-Order Index} is a multidimensional index that orders points by their \emph{Z-value}~\cite{zvalue}; contiguous chunks are grouped into pages.
    Given a query, the index finds the smallest and largest Z-value contained in the query rectangle and iterates through each page with Z-values in this range. Pages maintain min/max metadata per dimension, which allows queries to skip over irrelevant pages.
    \item The \emph{Hyperoctree}~\cite{octree} recursively subdivides space equally into hyperoctants (the $d$-dimensional analog to 2-dimensional quadrants), until the number of points in each leaf is below a predefined but tunable page size.
    \item The \emph{k-d tree}~\cite{kdtree} recursively partitions space using the median value along each dimension, until the number of points in each leaf falls below the page size. The dimensions are selected in a round robin fashion, in order of selectivity.
    \item \emph{Flood}, introduced in \Section{flood}. We use the implementation of \cite{nathan2020flood} with two changes: we use \tsunami's cost model instead of Flood's original random-forest-based cost model, and we perform refinement using binary search instead of learned per-cell models (see~\cite{nathan2020flood} for details). We verified that these changes did not meaningfully impact performance. Furthermore, removing per-cell models dramatically reduces Flood's index size (on average by 20$\times$~\cite{nathan2020flood}), and this allows us to more directly evaluate the impact of design differences between Flood and \tsunami without any confounding effects from implementation differences.
\end{CompactEnumerate}
There are a number of other multi-dimensional indexing techniques, such as Grid Files~\cite{gridfile}, UB-tree~\cite{ubtree}, and R$^*$-Tree~\cite{rstar-tree}. We decided not to evaluate against these because Flood already showed consistent superiority over them~\cite{nathan2020flood}. We also do not evaluate against other learned multi-dimensional indexes because they are either optimized for disk~\cite{qdtree-sigmod,lisa-sigmod} or optimize only based on the data distribution, not the query workload~\cite{zm-index,ml-index} (see \Section{related}).

\subsection{Datasets and Workloads}
\begin{table}[]
\small
\begin{tabular}{lllll}
\toprule
      & \textbf{TPC-H}. & \textbf{Taxi} & \textbf{Perfmon} & \textbf{Stocks}  \\
\midrule
\textbf{records}  & 300M & 184M & 236M & 210M \\
\textbf{query types}  & 5 & 6 & 5 & 5 \\
\textbf{dimensions}  & 8 & 9 & 7 & 7 \\
\textbf{size (GB)}  & 19.2 & 13.2 & 13.2 & 11.8 \\
\bottomrule
\end{tabular}
\centering
\caption{Dataset and query characteristics.}
\label{tab:data_params}
\vspace{-1em}
\end{table}

We evaluate indexes on three real-world and one synthetic dataset, summarized in \Table{data_params}. Queries are synthesized for each dataset, and include a mix of range filters and equality filters. The queries for each dataset comes from a certain number of query types (\Section{clustering}), each of which answers a different analytics question, with 100 queries of each type. All queries perform a \texttt{\small COUNT} aggregation. Since all indexes must pay the same fixed cost of aggregation, performing different aggregations would not change the relative ordering of indexes in terms of query performance.

The \textbf{Taxi} dataset comes from records of yellow taxi trips in New York City in 2018 and 2019~\cite{nyc-taxi}. It includes fields capturing pick-up and drop-off dates/times, pick-up and drop-off locations, trip distances, itemized fares, and driver-reported passenger counts. Our queries answer questions such as ``How common were single-passenger trips between two particular parts of Manhattan?'' and ``What month of the past year saw the most short-distance trips?''. Queries display skew over time (more queries over recent data), passenger count (different query types about very low and very high passenger counts), and trip distance (more queries about very short trip distances). Query selectivity varies from $0.25\%$ to $3.9\%$, with an average of $1.3\%$.

The performance monitoring dataset \textbf{Perfmon} contains logs of all machines managed by a major US university over the course of a year. It includes fields capturing log time, machine name, CPU usages, and system load averages. Our queries answer questions such as ``When in the last month did a certain set of machines experience high load?''. Queries display skew over time (more queries over recent data) and CPU usage (more queries over high usage). Query selectivity varies from $0.50\%$ to $4.9\%$, with an average of $0.79\%$. The original dataset has 23.6M records, but we use a scaled dataset with 236M records.

The \textbf{Stocks} dataset consists of daily historical stock prices of over 6000 stocks from 1970 to 2018~\cite{stocks}. It includes fields capturing daily prices (open, close, adjusted close, low, and high), trading volume, and the date. Our queries answer questions such as ``Which stocks saw the lowest intra-day price change while trading at high volume?'' and ``What one-year span in the past decade saw the most stocks close in a certain price range?''. Queries display skew over time (more queries over recent data) and volume (different query types about very low and very high volume). Query selectivity is tightly concentrated around $0.5\%\pm 0.04\%$. The original dataset has 21M records, but we use a scaled dataset with 210M records.

Our last dataset is \textbf{TPC-H}~\cite{tpch}.
For our evaluation, we use only the fact table, \texttt{lineitem}, with 300M records (scale factor 50) and create queries by using filters commonly found in the TPC-H query workload. Our queries include filters over quantity, extended price, discount, tax, ship mode, ship date, commit date, and receipt date. They answer questions such as ``How many high-priced orders in the past year used a significant discount?'' and ``How many shipments by air had below ten items?''. Query selectivity varies from $0.40\%$ to $0.64\%$, with an average of $0.54\%$.

\subsection{Overall Results}
\label{sec:overall_results}

\begin{figure*}[]
    \centering
    \subfloat{
        \includegraphics[width=0.24\textwidth,clip]{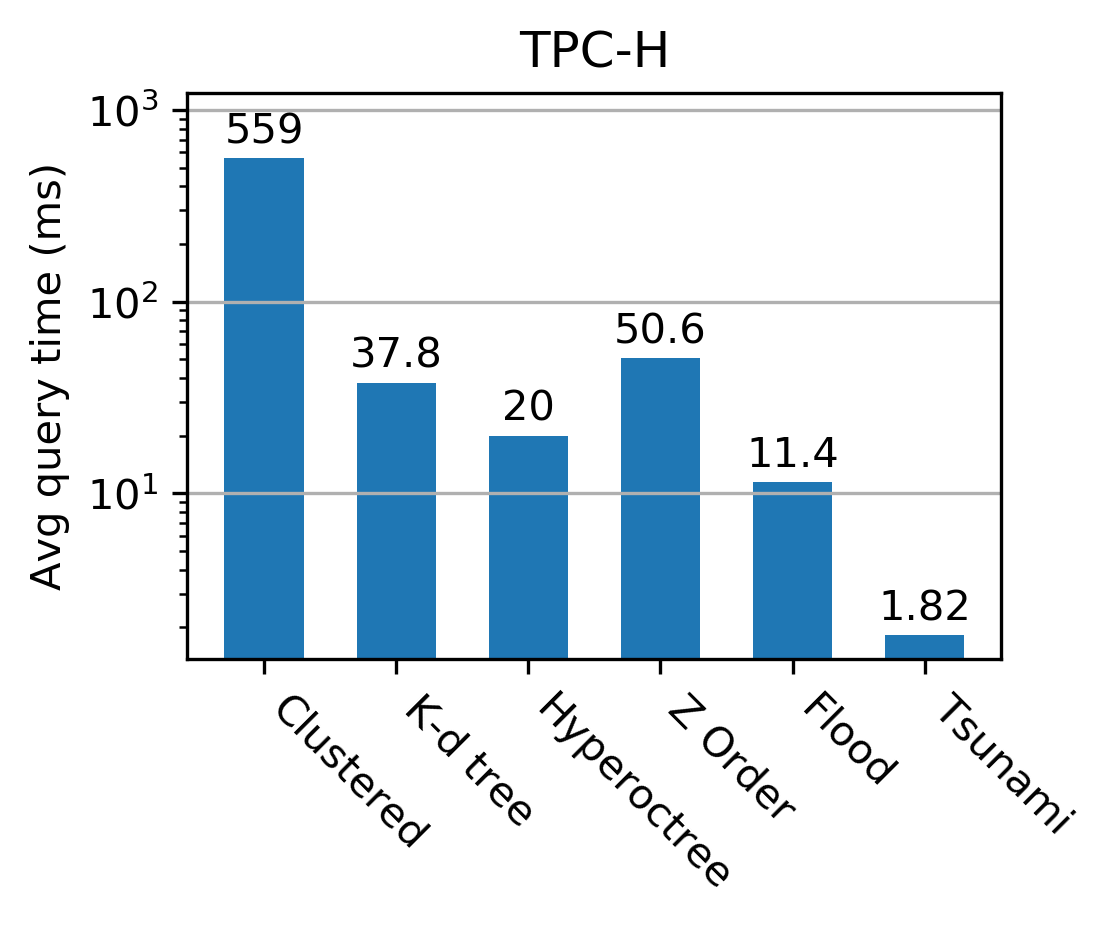}
    }
    ~
    \subfloat{
        \includegraphics[width=0.24\textwidth,clip]{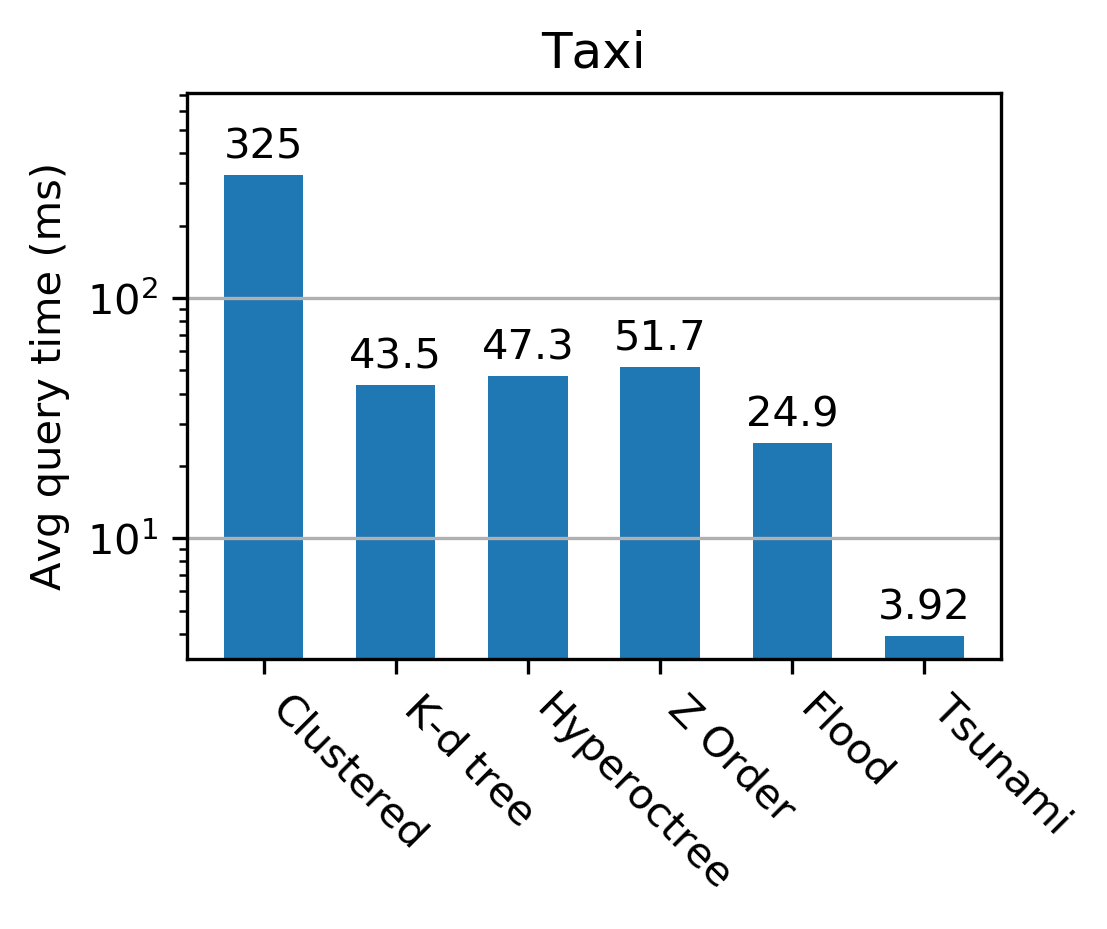}
    }
    ~
    \subfloat{
        \includegraphics[width=0.24\textwidth,clip]{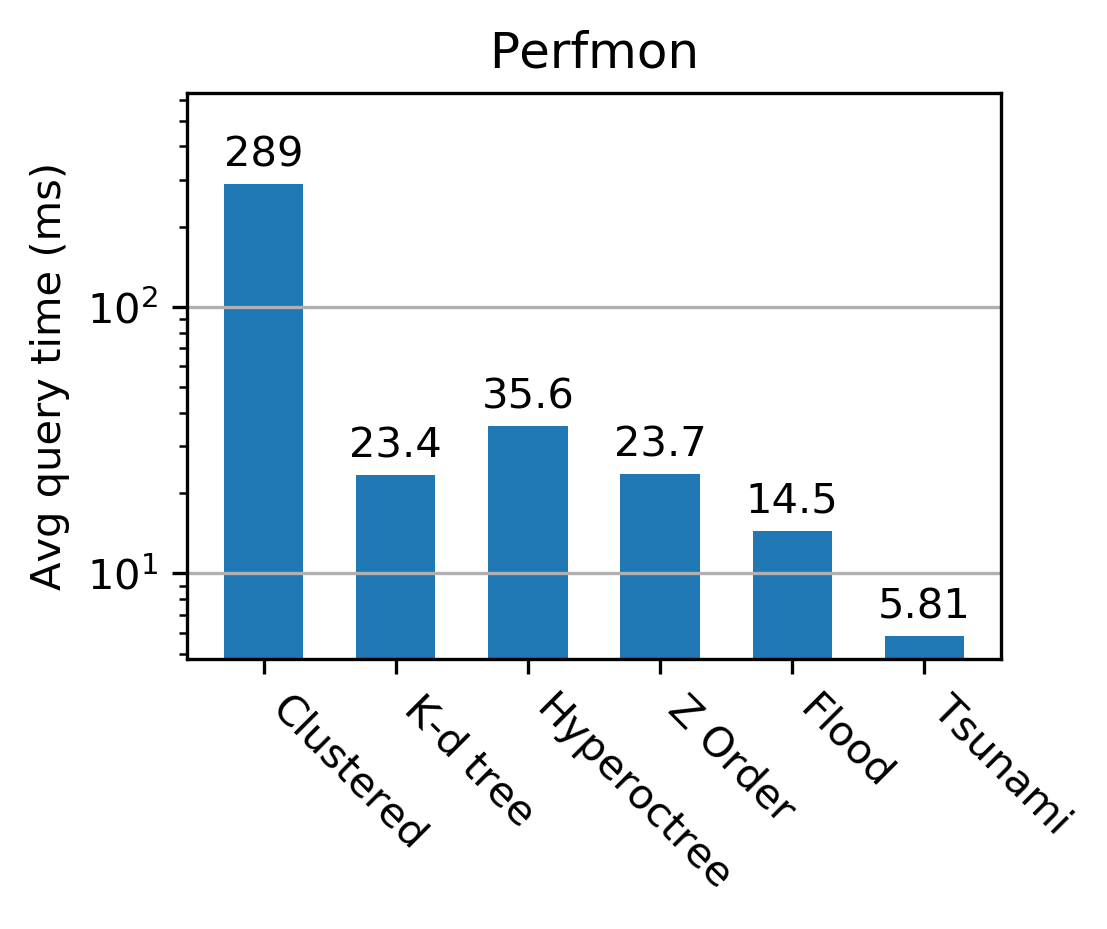}
    }
    ~
    \subfloat{
        \includegraphics[width=0.24\textwidth,clip]{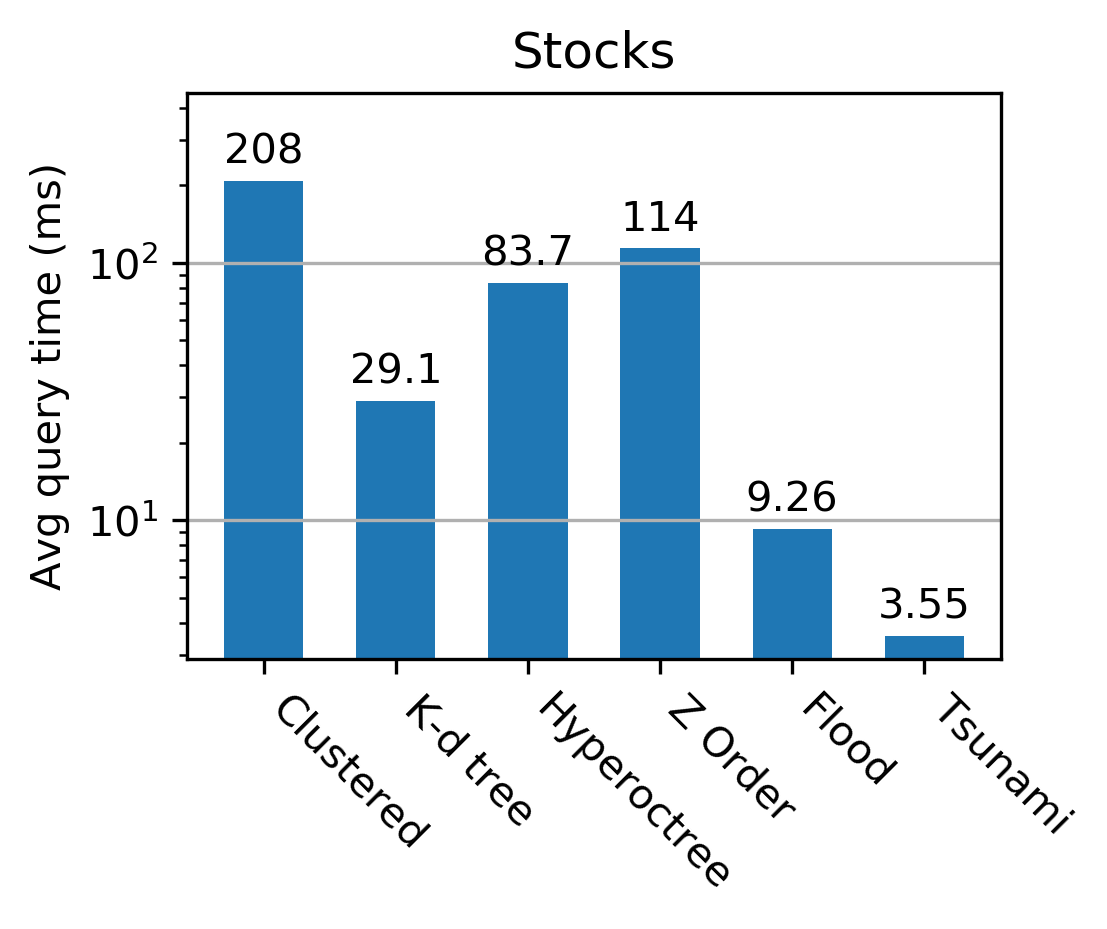}
    }
    \caption{\tsunami achieves up to 6$\times$ faster queries than Flood and up to 11$\times$ faster queries than the fastest non-learned index.}
    \label{fig:punchline}
    \vspace{-1em}
\end{figure*}

\begin{figure*}[]
    \centering
    \includegraphics[width=0.7\textwidth]{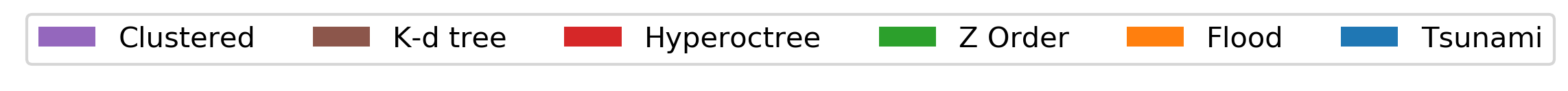}
    \\
    \vspace{-1em}
    \subfloat{
        \includegraphics[width=0.24\textwidth,clip]{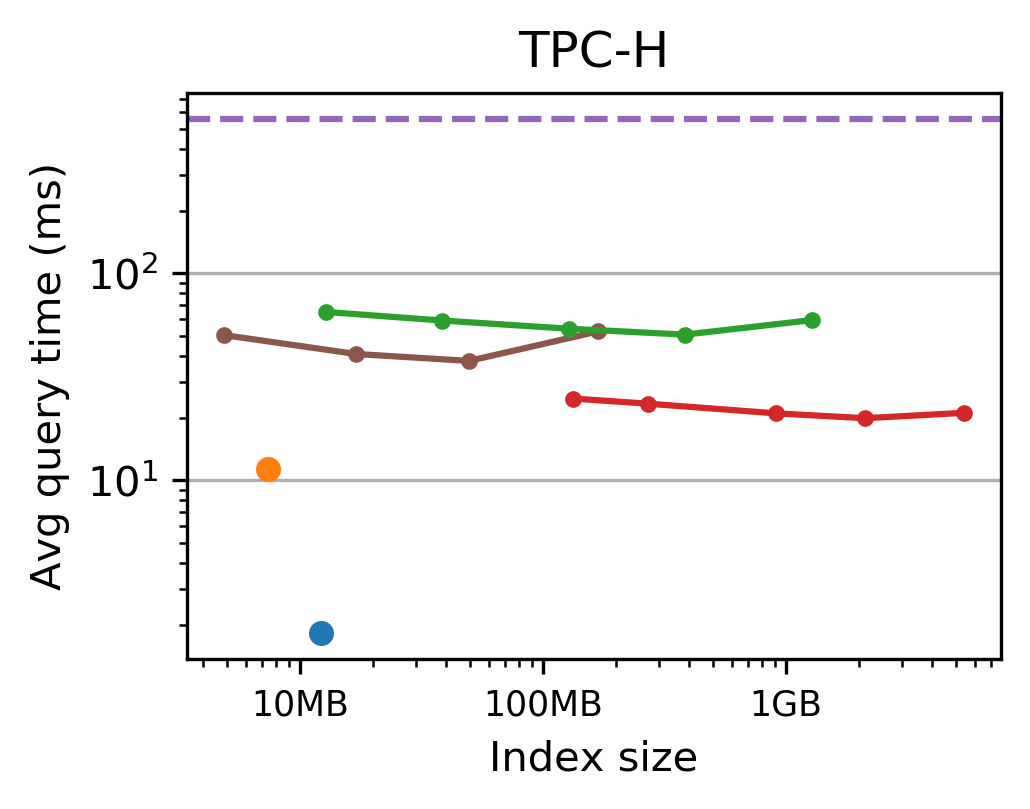}
    }
    ~
    \subfloat{
        \includegraphics[width=0.24\textwidth,clip]{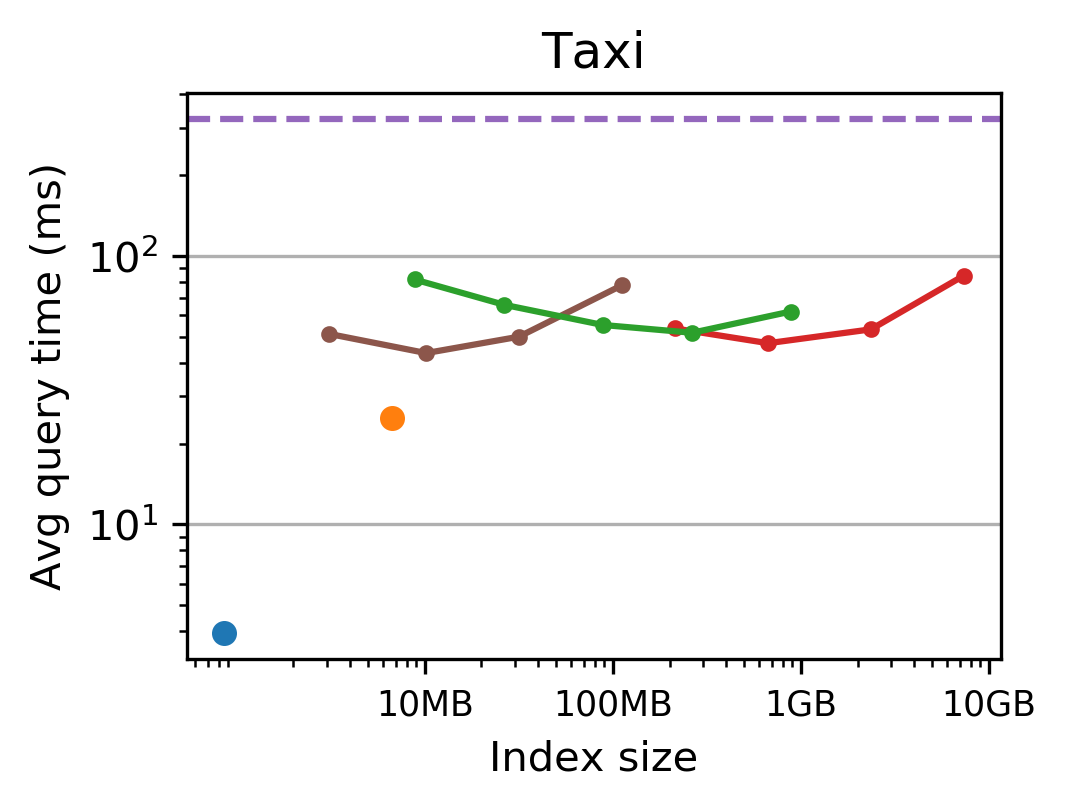}
    }
    ~
    \subfloat{
        \includegraphics[width=0.24\textwidth,clip]{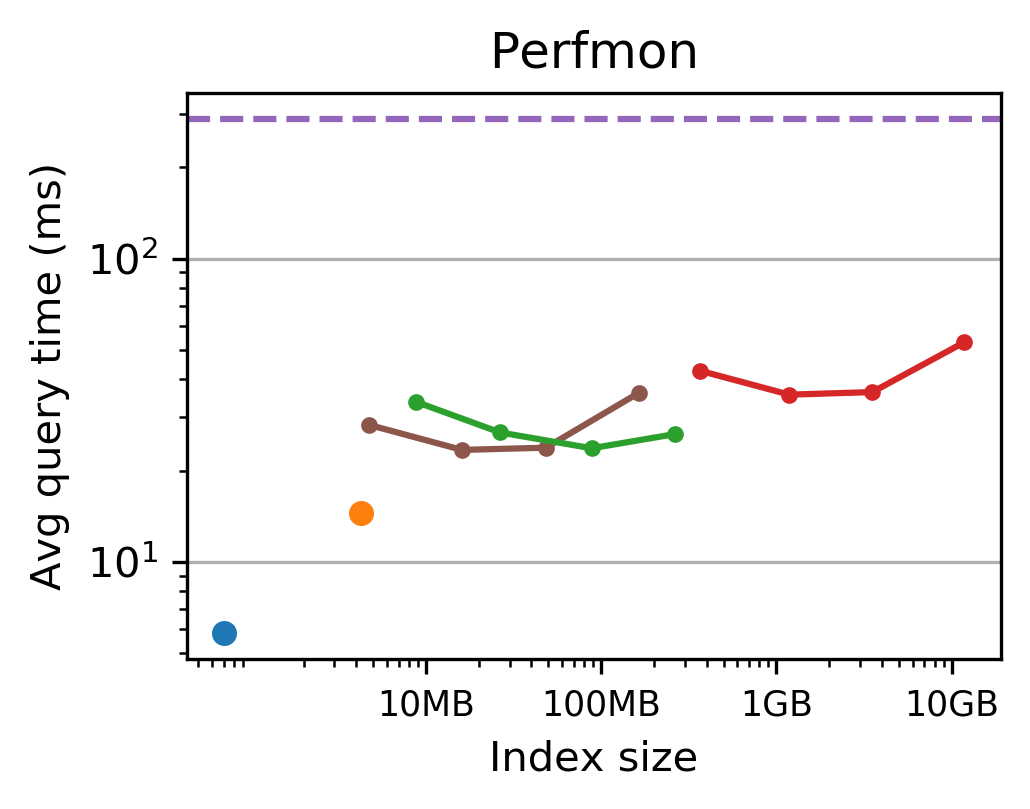}
    }
    ~
    \subfloat{
        \includegraphics[width=0.24\textwidth,clip]{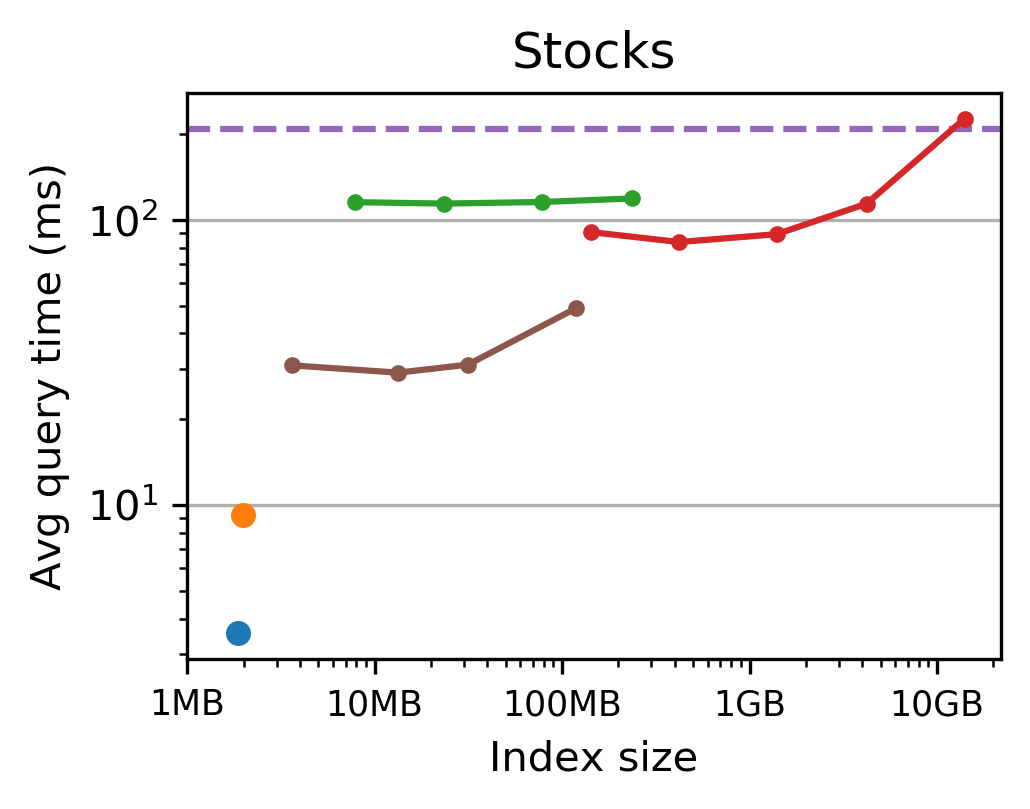}
    }
    \caption{\tsunami uses up to 8$\times$ less memory than Flood and 7-170$\times$ less memory than the fastest tuned non-learned index.}
    \label{fig:pareto}
    \vspace{-1em}
\end{figure*}

\Figure{punchline} compares \tsunami to Flood and the non-learned baselines. \tsunami and Flood are automatically optimized for each dataset/workload. For the non-learned baselines, we tuned the page size to achieve best performance on each dataset/workload. \tsunami is consistently the fastest of all the indexes across datasets and workloads, and achieves up to 6$\times$ faster queries than Flood and up to 11$\times$ faster queries than the fastest non-learned index.

\Table{tsunami_stats} shows statistics of the optimized \tsunami index structure. The \gridtree depth and the number of leaf regions are relatively low, which confirms that the \gridtree is lightweight, as desired. Because skew does not occur uniformly across data space, the number of points in each region can vary by over an order of magnitude.

\begin{table}[]
\small
\begin{tabular}{lllll}
\toprule
      & \textbf{TPC-H} & \textbf{Taxi} & \textbf{Perfmon} & \textbf{Stocks}  \\
\midrule
\textbf{\textit{Tsunami}} \\
\cmidrule(r){1-1}
\textbf{Num \gridtree nodes}& 39    & 35    & 42    & 54 \\
\textbf{\gridtree depth}    & 4     & 2     & 4     & 4 \\
\textbf{Num leaf regions}   & 27    & 31     & 36    & 39 \\
\textbf{Min points per region}      & 3.5M  & 1.9M  & 2.6M  & 2.4M \\
\textbf{Median points per region}   & 5.9M  & 3.3M  & 3.7M  & 3.2M \\
\textbf{Max points per region}      & 10M   & 6.7M  & 26M   & 41M \\
\textbf{Avg FMs per region}   & 0.67  & 0.55  & 0     & 1.1 \\
\textbf{Avg CCDFs per region}  & 1.3   & 1.9   & 1.75  & 1.8 \\
\textbf{Total num grid cells}  & 1.5M   & 99K   & 80K  & 220K \\
\midrule
\textbf{\textit{Flood}} \\
\cmidrule(r){1-1}
\textbf{Num grid cells}  & 920K   & 840K & 530K  & 250K \\
\bottomrule
\end{tabular}
\centering
\caption{Index Statistics after Optimization.}
\label{tab:tsunami_stats}
\vspace{-1em}
\end{table}

The \gridtree typically has a low number of nodes (\Table{tsunami_stats}), so the vast majority of \tsunami's index size comes from the cell lookup tables for the \augmentedgrids in each region. \tsunami often has fewer total grid cells than Flood (\Table{tsunami_stats}) because partitioning space via the \gridtree gives \tsunami fine-grained control over the number of cells to allocate in each region, whereas Flood must often over-provision partitions to deal with query skew (see \Section{query_skew_challenges}). \Figure{pareto} shows that as a result of having fewer cells, \tsunami uses up to 8$\times$ less memory than Flood. Furthermore, \tsunami is between 7$\times$ to 170$\times$ smaller than the fastest optimally-tuned non-learned index across the four datasets.

\subsection{Adaptibility}
\label{sec:adaptibility_results}
\begin{figure}[]
    \centering
    \includegraphics[width=\columnwidth]{figures/legend_pareto.png}
    \\
    \vspace{-1em}
    \subfloat{
        \includegraphics[width=0.45\columnwidth,clip]{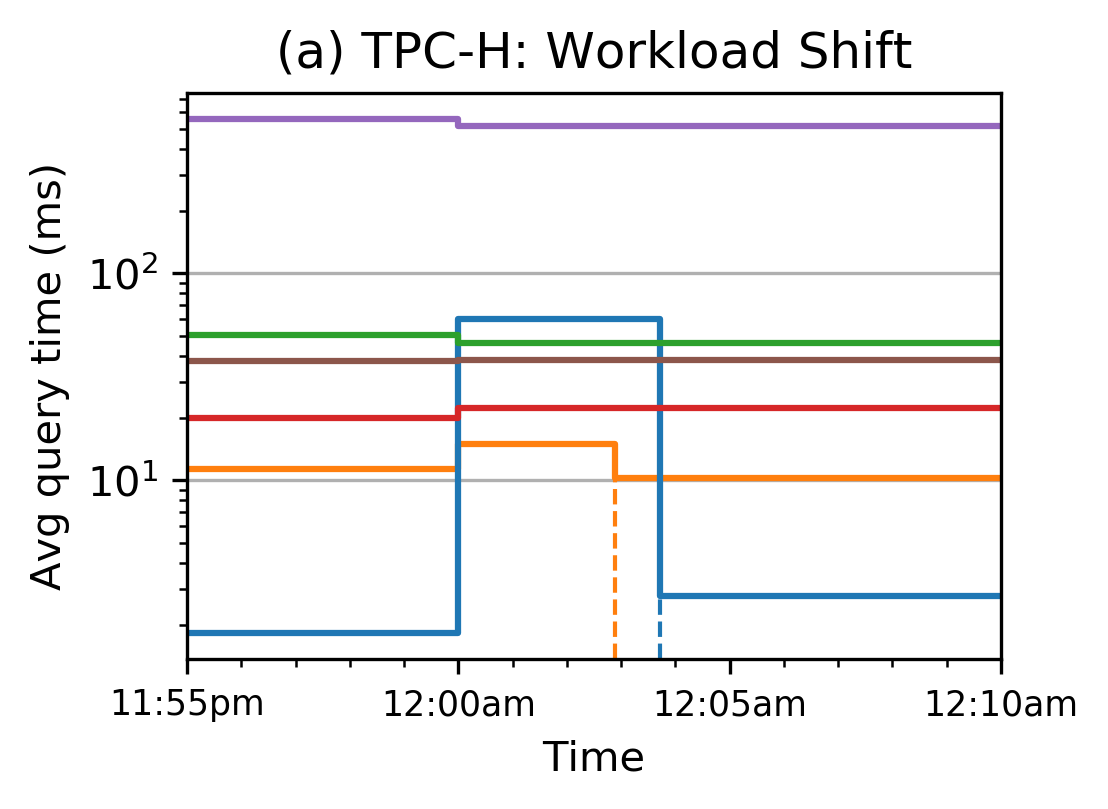}
        \label{fig:workload_shift}
    }
    ~
    \subfloat{
        \includegraphics[width=0.51\columnwidth,clip]{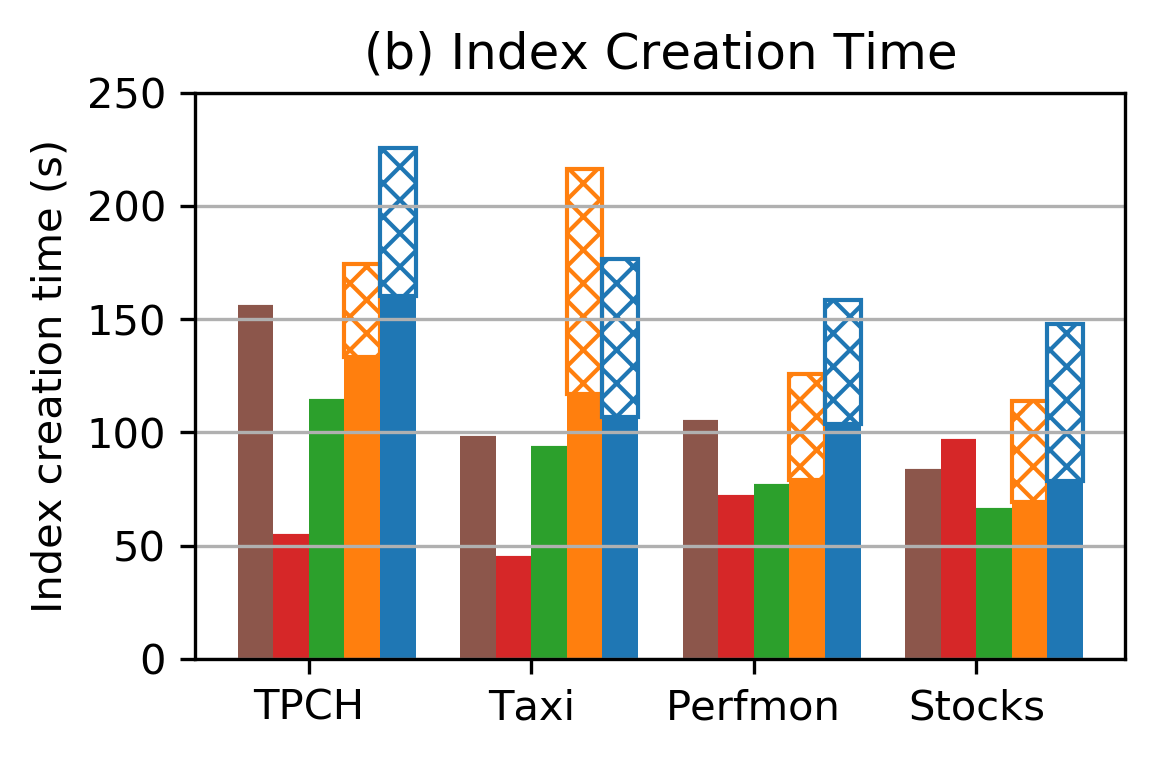}
        \label{fig:index_creation}
    }
    \caption{(a) After the query workload changes at midnight, \tsunami re-optimizes and re-organizes within 4 minutes to maintain high performance. (b) Comparison of index creation times (solid bars = data sorting time, hatched bars = optimization time).}
    \label{fig:adaptibility}
    \vspace{-1em}
\end{figure}

\tsunami is able to quickly adapt to changes in the query workload by re-optimizing its layout for the new query workload and re-organizing the data based on the new layout. In \Figure{workload_shift}, we simulate a scenario in which the query workload over the TPC-H dataset changes at midnight: the original query workload is replaced by a new workload with queries drawn from five new query types. This causes performance on the learned indexes to degrade. \tsunami (as well as Flood) automatically detects the workload shift (see \Section{future}) and triggers a re-optimization of the index layout for the new query workload. \tsunami's re-optimization and data re-organization over 300M rows finish within 4 minutes, and its high query performance is restored. This shows that \tsunami is highly adaptive for scenarios in which the data or workload changes infrequently (e.g., every day). The non-learned indexes are not re-tuned after the workload shift, because in practical settings, it is unlikely that a database administrator will be able to manually tune the index for every workload change.

\Figure{index_creation} shows the index creation time in greater detail for \tsunami and the baselines. All indexes require time to sort the data based on the index layout, shown as solid bars. The learned approaches additionally require time to perform optimization based on the dataset and query workload, shown as the hatched bars. Even for the largest datasets, the entire index creation time for \tsunami remains below 4 minutes.

\subsection{Scalability}
\label{sec:scalability_results}
Throughout this subsection, \tsunami and Flood are re-optimized for each dataset/workload configuration, while the non-learned indexes use the same page size and dimension ordering as they were tuned for the full TPC-H dataset/workload in \Section{overall_results}.

\NewPara{Number of Dimensions.}
To show how \tsunami scales with dimensions, and how correlation affects scalability, we create two groups of synthetic $d$-dimensional datasets with 100M records. Within each group, datasets vary by number of dimensions ($d\in\{4, 8, 12, 16, 20\}$). Datasets in the first group show no correlation and points are sampled from i.i.d. uniform distributions. For datasets in the second group, half of the dimensions have uniformly sampled values, and dimensions in the other half are linearly correlated to dimensions in the first half, either strongly ($\pm1\%$ error) or loosely ($\pm10\%$ error). For each dataset, we create a query workload with four query types. Earlier dimensions are filtered with exponentially higher selectivity than later dimensions, and queries are skewed over the first four dimensions.

\begin{figure}[]
    \centering
    \includegraphics[width=\columnwidth]{figures/legend_pareto.png}
    \\
    \vspace{-1em}
    \subfloat{
        \includegraphics[width=0.48\columnwidth,clip]{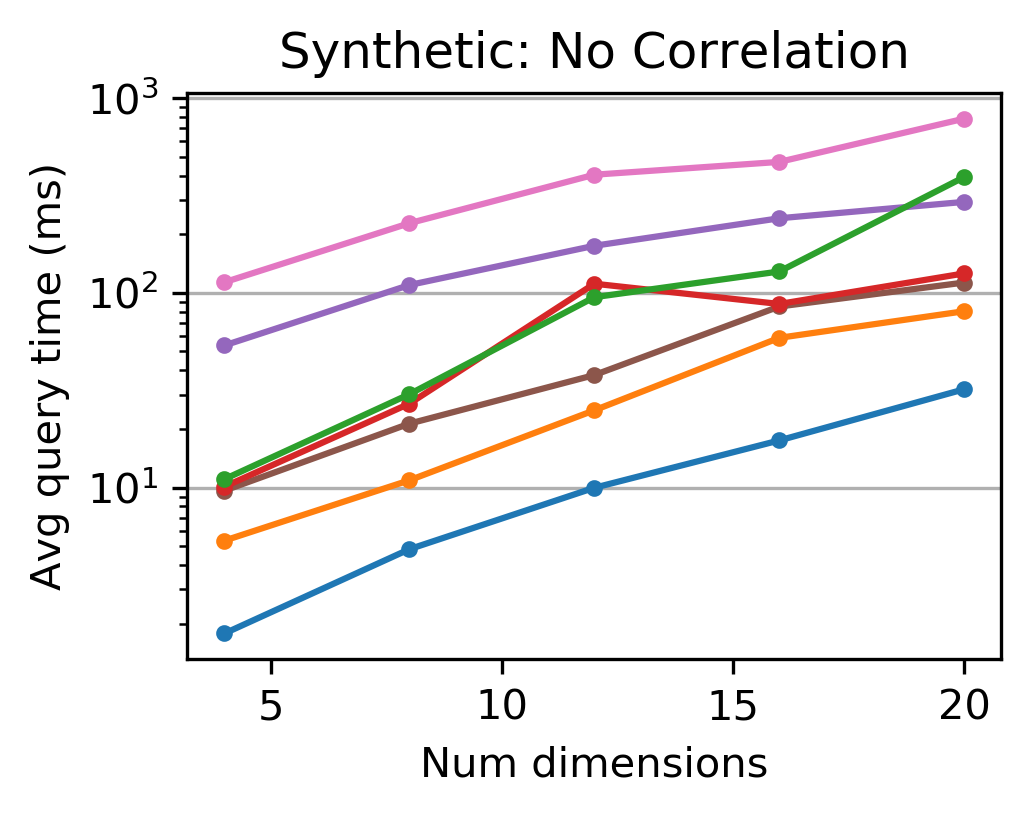}
    }
    ~
    \subfloat{
        \includegraphics[width=0.48\columnwidth,clip]{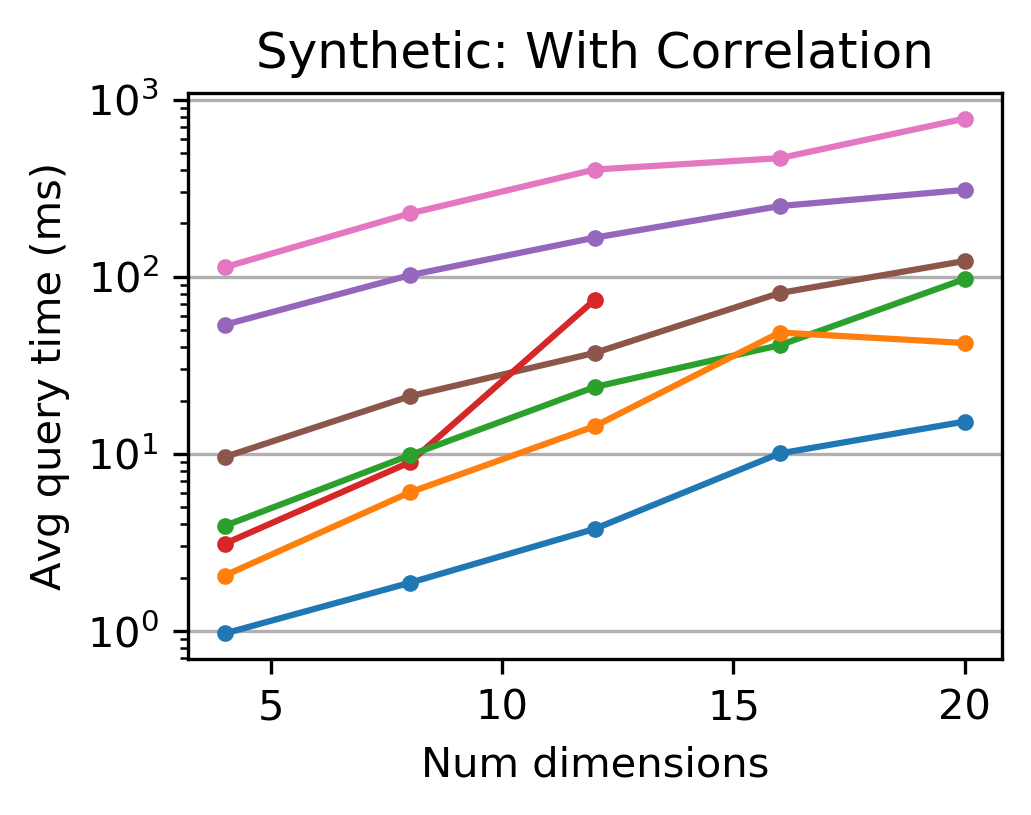}
    }
    \caption{\tsunami continues outperform other indexes at higher dimensions. In particular, \augmentedgrid helps \tsunami delay the curse of dimensionality on correlated data.}
    \label{fig:vary_num_dimensions}
    \vspace{-1em}
\end{figure}

\Figure{vary_num_dimensions} shows that in both cases, \tsunami continues to outperform the other indexes at higher dimensions. In particular, the \augmentedgrid is able to take advantage of correlations to effectively reduce the dimensionality of the dataset. This helps \tsunami delay the curse of dimensionality: \tsunami has around the same performance on each $d$-dimensional correlated dataset as it does on the $(d-4)$-dimensional uncorrelated dataset.

\begin{figure}[]
    \centering
    \includegraphics[width=\columnwidth]{figures/legend_pareto.png}
    \\
    \vspace{-1em}
    \subfloat{
        \includegraphics[width=0.48\columnwidth,clip]{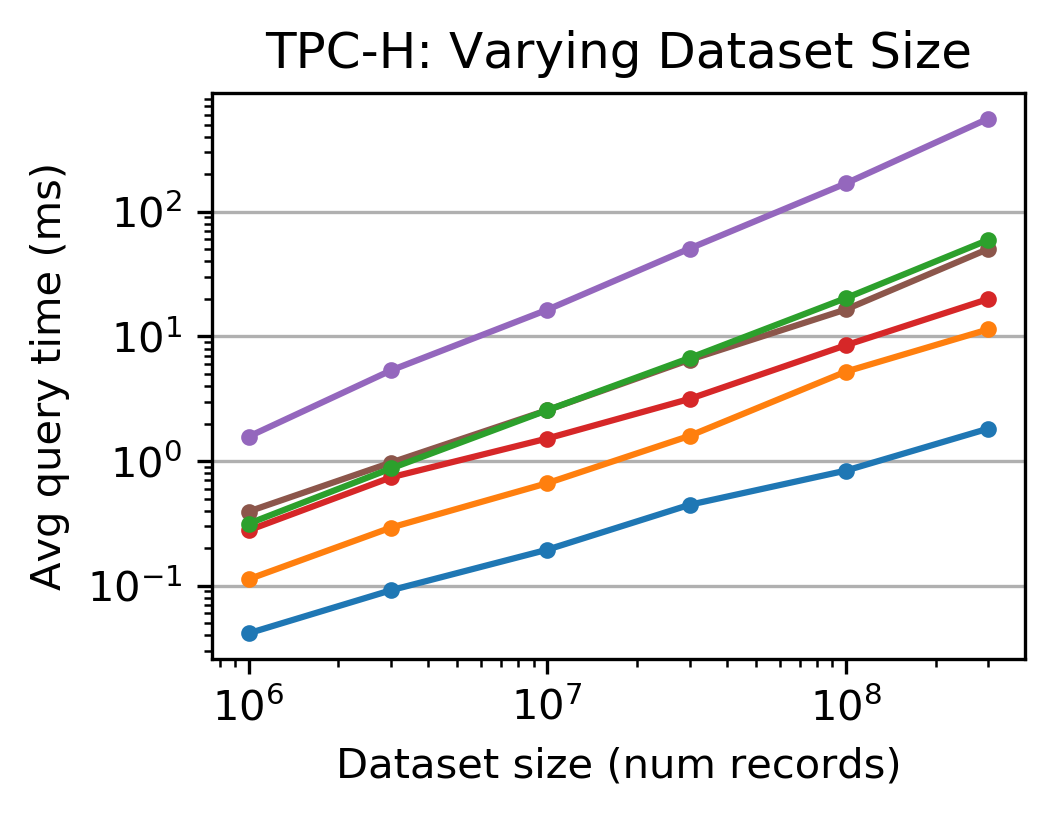}
        \label{fig:vary_size}
    }
    ~
    \subfloat{
        \includegraphics[width=0.48\columnwidth,clip]{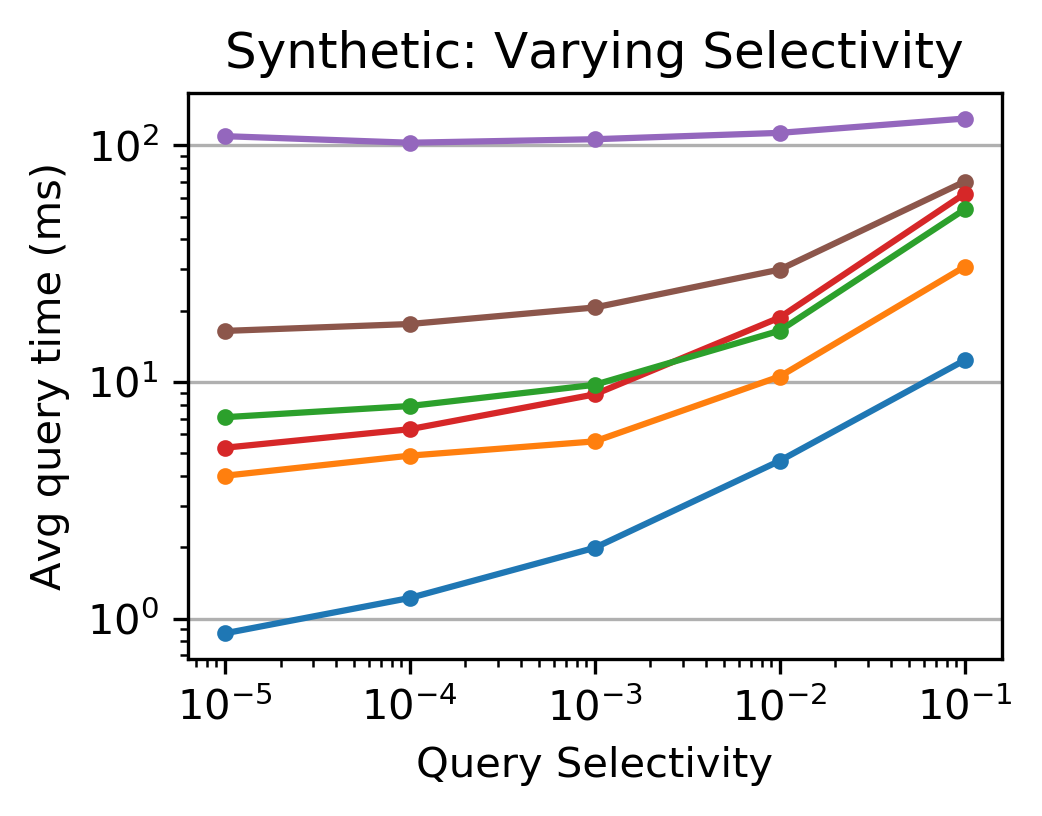}
        \label{fig:vary_selectivity}
    }
    \caption{\tsunami maintains high performance across dataset sizes and query selectivities.}
    \label{fig:vary_size_selectivity}
    \vspace{-1em}
\end{figure}

\NewPara{Dataset Size.}
To show how \tsunami scales with dataset size, we sample records from the TPC-H dataset to create smaller datasets. We run the same query workload as on the full dataset. \Figure{vary_size} shows that across dataset sizes, \tsunami maintains its performance advantage over Flood and non-learned indexes.

\NewPara{Query Selectivity.}
To show how \tsunami performs at different query selectivities, we use the 8-dimensional synthetic dataset/workload with correlation (explained above) and scale filter ranges up and down equally in each dimension in order to achieve between 0.001\% and 10\% selectivity. \Figure{vary_selectivity} shows that \tsunami performs well at all selectivities. The relative performance benefit of \tsunami is less apparent at 10\% selectivity because aggregation time becomes a bottleneck.

\subsection{Drill-down into Components}
\label{sec:components_results}
\begin{figure}[]
    \subfloat{
        \includegraphics[width=0.46\columnwidth,clip]{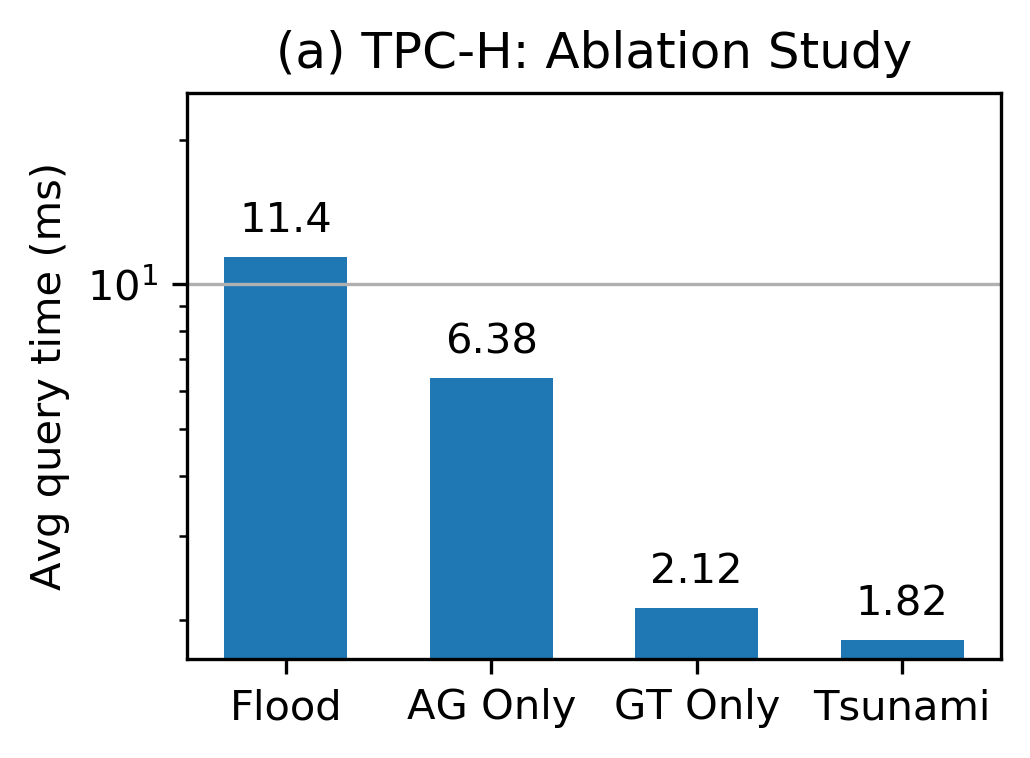}
        \label{fig:lesion}
    }
    ~
    \subfloat{
        \includegraphics[width=0.50\columnwidth,clip]{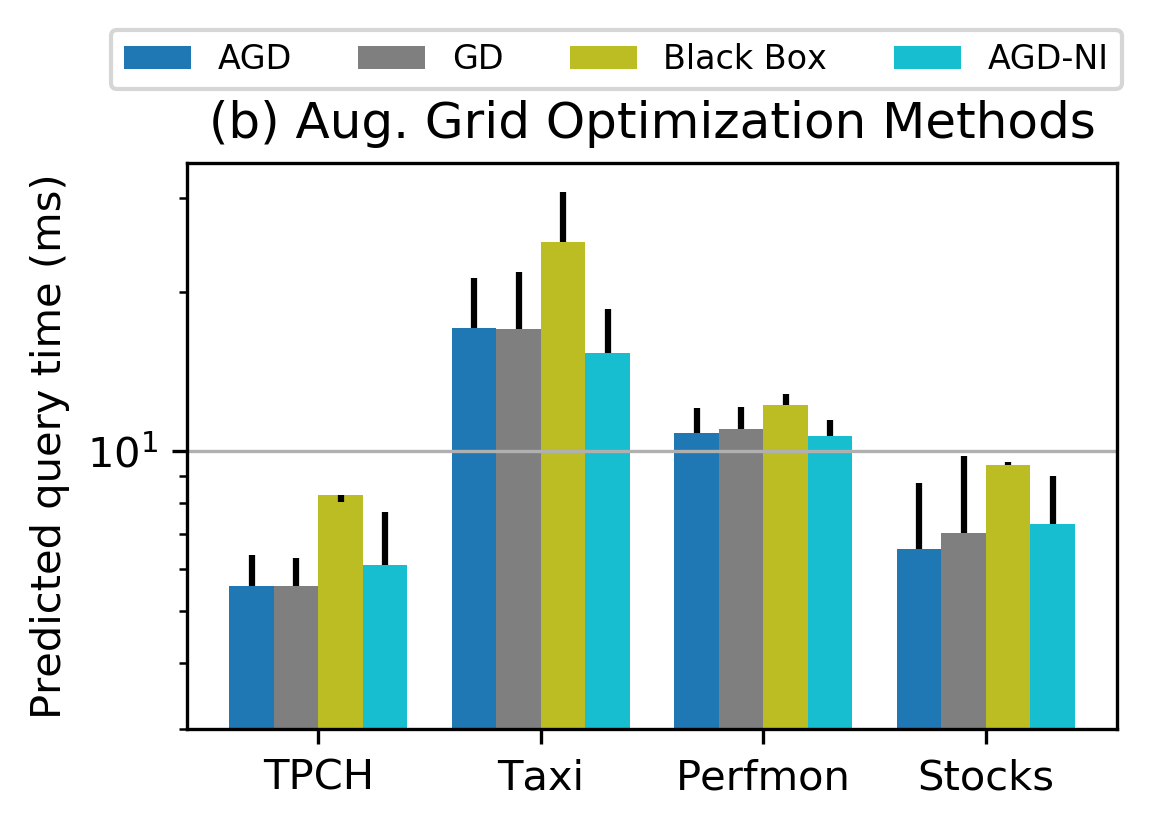}
        \label{fig:compare_opt}
    }
    \caption{(a) \augmentedgrid and \gridtree both contribute to \tsunami's performance. (b) Comparison of optimization methods. Bars show the predicted query time according to our cost model. Error bars show the actual query time.}
    \label{fig:drilldown}
    \vspace{-1em}
\end{figure}

\Figure{lesion} shows the relative performance of only using the \augmentedgrid (i.e., one \augmentedgrid over the entire data space) and of only using \gridtree (i.e., with an instantiation of Flood in each leaf region).
\gridtree contributes the most to \tsunami's performance, but \augmentedgrid also boosts performance significantly over Flood. \gridtree-only performs almost as well as \tsunami because partitioning data space via the \gridtree often already has the unintentional but useful side effect of mitigating data correlations.

We now evaluate \augmentedgrid's optimization procedure, which can be broken into two independent parts: the accuracy of the cost model (\Section{cost_model}) and the ability of Adaptive Gradient Descent (\Section{agd}) to minimize cost (i.e., average query time, predicted by the cost model). For each of our four datasets/workloads, we run Adaptive Gradient Descent (AGD) to find a low-cost \augmentedgrid over the entire data space. We compare with three alternative optimization methods, all using the same cost model:
\begin{CompactEnumerate}
    \item \textit{Gradient Descent (GD)} uses the same initial $(S_0,P_0)$ as AGD, then performs gradient descent over $P$, without ever changing the skeleton.
    \item \textit{Black Box} starts with the same initial $(S_0,P_0)$ as AGD, then optimizes $S$ and $P$ according to the basin hopping algorithm, implemented in SciPy~\cite{scipy}, for 50 iterations.
    \item \textit{Adaptive Gradient Descent with naive initialization (AGD-NI)} sets the initial skeleton $S_0$ to use $CDF(X)$ for each dimension, then runs AGD.
\end{CompactEnumerate}
\Figure{compare_opt} shows the lowest cost achieved by each optimization method. There are several insights. First, Black Box performs worse than the gradient descent variants, which implies that using domain knowledge and heuristics to guide the search process provides an advantage. Second, Adaptive Gradient Descent usually achieves only marginally better predicted query time than Gradient Descent, which implies that for our tested datasets, our heuristics created a good initial skeleton $S_0$. Third, Adaptive Gradient Descent is able to find a low-cost grid even when starting from a naive skeleton, which implies that the local search over skeletons is able to effectively switch to better skeletons. For the Taxi dataset, AGD-NI is even able to find a lower-cost configuration than AGD.

\Figure{compare_opt} additionally shows the error between the predicted query time using the cost model and the actual query time when running the queries of the workload. The average error of the model for all optimized configurations shown in \Figure{compare_opt} is only 15\%.
\section{Related Work}
\label{sec:related}
\NewPara{Traditional Multi-dimensional Indexes.}
There is a rich corpus of work dedicated to multi-dimensional indexes, and many commercial database systems have turned to multi-dimensional indexing schemes. 
For example, Amazon Redshift organizes points by Z-order~\cite{z-order}, which maps multi-dimensional points onto a single dimension for sorting~\cite{redshift, oracle-zorder, amazon-zorder}. With spatial dimensions, SQL Server allows Z-ordering~\cite{sql-server}, and IBM Informix uses an R-Tree~\cite{ibm-rtree}. 
Other multi-dimensional indexes include K-d trees, octrees, R$^*$ trees, UB trees (which also make use of the Z-order), and Grid Files~\cite{gridfile}, among many others (see \cite{Ooi_indexingspatial,spatialindexsurvey} for a survey).
There has also been work on automatic index selection~\cite{self-driving, self-driving2, self-driving3}. 
However, these approaches mainly focus on creating secondary indexes, whereas \tsunami co-optimizes the index and data storage.

\NewPara{Learned Indexes.} Recent work by Kraska et al.~\cite{rmi} proposed the idea of replacing traditional database indexes with learned models that predict the location of a key in a dataset. Their learned index, called the Recursive Model Index (RMI), and various improvements on the RMI~\cite{alex,pgm-index,radixspline,fiting-tree,xindex}, only handle one-dimensional keys.

Since then, there has been a corpus of work on extending the ideas of the learned index to spatial and multi-dimensional data. The most relevant work is Flood~\cite{nathan2020flood}, described in \Section{flood}. Learning has also been applied to the challenge of reducing I/O cost for disk-based multi-dimensional indexes. Qd-tree~\cite{qdtree-sigmod} uses reinforcement learning to construct a partitioning strategy that minimizes the number of disk-based blocks accessed by a query. LISA~\cite{lisa-sigmod} is a disk-based learned spatial index that achieves low storage consumption and I/O cost while supporting range queries, nearest neighbor queries, and insertions and deletions. \tsunami and these works share the idea that a multi-dimensional index can be instance-optimized for a particular use case by learning from the dataset and query workload.

Past work has also aimed to improve traditional indexing techniques by learning the data distribution. The ZM-index~\cite{zm-index} combines the standard Z-order space-filling curve~\cite{z-order} with the RMI from~\cite{rmi} by mapping multi-dimensional values into a single-dimensional space, which is then learnable using models. The ML-index~\cite{ml-index} combines the ideas of iDistance~\cite{idistance} and the RMI to support range and KNN queries. Unlike \tsunami, these works only learn from the data distribution, not from the query workload.

\NewPara{Data Correlations.} There is a body of work on discovering and taking advantage of column correlations. BHUNT~\cite{bhunt}, CORDS~\cite{cords}, and Pyro~\cite{pyro} automatically discover algebraic constraints, soft functional dependencies, and approximate dependencies between columns, respectively. CORADD~\cite{coradd} recommends materialized views and indexes based on correlations. Correlation Map~\cite{correlationmaps} aims to reduce the size of B+Tree secondary indexes by creating a mapping between correlated dimensions. Hermit~\cite{hermit} is a learned secondary index that achieves low space usage by capturing monotonic correlations and outliers between dimensions. Although the functional mappings in the \augmentedgrid are conceptually similar to Correlation Map and Hermit, our work is more focused on how to incorporate correlation-aware techniques into a multi-dimensional index.

\NewPara{Query Skew.} The existence of query skew has been extensively reported in settings where data is accessed via single-dimensional keys (i.e., ``hot keys'')~\cite{ycsb,facebook-workload,facebook-rocksdb}. In particular, key-value store workloads at Facebook display strong key-space locality: hot keys are closely located in the key space~\cite{facebook-rocksdb}. Instead of relying on caches to reduce query time for frequently accessed keys, \tsunami automatically partitions data space using the \gridtree to account for query skew.

\section{Future Work}
\label{sec:future}

\NewPara{Complex Correlations.} Although \augmentedgrid introduces techniques to address data correlations, there are more opportunities on the table. First, functional mappings are not robust to outliers: one outlier can significantly increase the error bound of the mapping. We can address this by placing outliers in a separate buffer, similar to Hermit~\cite{hermit}. Second, \augmentedgrid might not efficiently capture more complex correlation patterns, such as temporal/periodic patterns and correlations due to functional dependencies over more than two dimensions. To handle these correlations, we intend to introduce new correlation-aware partitioning strategies to the \augmentedgrid.

\NewPara{Categorical dimensions.} 
Values of categorical dimensions typically have no semantically meaningful sort order, so they are sorted alphanumerically by default. However, we can improve performance by imposing our own sort order. In the \augmentedgrid, we can sort categorical values by co-access frequency: values that are commonly accessed together in the same query should ideally be placed in the same grid partition, so that a query that accesses them needs to scan fewer partitions and points.

\NewPara{Data and Workload Shift.}
\tsunami can quickly adapt to workload changes but does not currently have a way to detect when the workload characteristics have changed sufficiently to merit re-optimization. To do this, \tsunami could detect when an existing query type (\Section{clustering}) disappears, a new query type appears, or when the relative frequencies of query types change. \tsunami could also detect when the query skew of a particular \gridtree region has deviated from its skew after the initial optimization. Additionally, \tsunami is completely re-optimized for each new workload. However, \tsunami could be incrementally adjusted, e.g. by only re-optimizing the \augmentedgrids whose regions saw the most significant workload shift.

\tsunami currently only supports read-only workloads. To support insertions, each leaf node in the \gridtree could maintain a sibling node that acts as a delta index~\cite{delta-index} in which updates are buffered and periodically merged into the main node.

\NewPara{Persistence} \tsunami's techniques for reducing query skew and handling correlations are not restricted to in-memory scenarios and could be incorporated into a multi-dimensional index for data resident on disk or SSD, perhaps by combining ideas from qd-tree~\cite{qdtree-sigmod} or LISA~\cite{lisa-sigmod}.
\section{Conclusion}
\label{sec:conclusion}
Recent work has introduced the idea of learned multi-dimensional indexes, which outperform traditional multi-dimensional indexes by co-optimizing the index layout and data storage for a particular dataset and query workload. We design \tsunami, a new in-memory learned multi-dimensional index that pushes the boundaries of performance by automatically adapting to data correlations and query skew. \tsunami introduces two modular data structures---\gridtree and \augmentedgrid---that allow it to outperform existing learned multi-dimensional indexes by up to 6$\times$ in query throughput and 8$\times$ in space. Our results take us one step closer towards a robust learned multi-dimensional index that can serve as a building block in larger in-memory database systems.

\NewPara{Acknowledgements.}
This research is supported by Google, Intel, and Microsoft as part of the MIT Data Systems and AI Lab (DSAIL) at MIT, NSF IIS 1900933, DARPA Award 16-43-D3M-FP040, and the MIT Air Force Artificial Intelligence Innovation Accelerator (AIIA). Research was sponsored by the United States Air Force Research Laboratory and was accomplished under Cooperative Agreement Number FA8750-19-2-1000. The views and conclusions contained in this document are those of the authors and should not be interpreted as representing the official policies, either expressed or implied, of the United States Air Force or the U.S. Government. The U.S. Government is authorized to reproduce and distribute reprints for Government purposes notwithstanding any copyright notation herein.


\bibliographystyle{ACM-Reference-Format}
\bibliography{main}

\end{document}